\documentclass{statsoc}
\usepackage[english]{babel}
\usepackage[a4paper]{geometry}
\usepackage{bm}
\usepackage{amsmath}
\usepackage{amssymb}
\usepackage{natbib}
\usepackage{relsize}
\usepackage[table]{xcolor}
\usepackage{multirow}

\usepackage{lscape}
\usepackage{etoolbox}
\usepackage{url}
\DeclareGraphicsExtensions{.pdf,.png,.jpg,.eps}

\makeatletter
\patchcmd{\@makecaption}
  {\parbox}
  {\advance\@tempdima-\fontdimen2} 
  {}{}
\makeatother  

\usepackage{graphicx}
\usepackage[caption=false]{subfig}
\usepackage{enumerate}
\usepackage{comment}

\newtheorem{thm}{Theorem}
\newtheorem{corl}{Corollary}
\title[MDD Priors]{\bf{Mixture Data-Dependent Priors }}
\author[Author 1 {\it et al.}]{Leonardo Egidi}
\address{Dipartimento di Scienze Statistiche, Universit\`a degli Studi di Padova, 
Padova,
Italy.}
\email{egidi@stat.unipd.it}
\author[Author 2]{Francesco Pauli}
\address{Dipartimento di Scienze Economiche, Aziendali, Matematiche e Statistiche `Bruno de Finetti', Universit\`{a} degli Studi di Trieste, Italy}
\email{francesco.pauli@deams.units.it}
\author[Egidi Pauli Torelli]{Nicola Torelli}
\address{Dipartimento di Scienze Economiche, Aziendali, Matematiche e Statistiche `Bruno de Finetti', Universit\`{a} degli Studi di Trieste, Italy}
\email{nicola.torelli@deams.units.it}

\begin{document}

\begin{abstract}

 We propose a two-component mixture of a noninformative (diffuse) and an informative prior distribution, weighted through the data in such a way to prefer the first component if a prior-data conflict arises. The data-driven approach for computing the mixture weights makes this class data-dependent. Although rarely used with any theoretical motivation, data-dependent priors are often used for different reasons, and their use has been a lot debated over the last decades. However, our approach is justified in terms of Bayesian inference as an approximation of a hierarchical model and as a conditioning on a data statistic. This class of priors turns out to provide less information than an informative prior, perhaps it represents a suitable option for not dominating the inference in presence of small samples. First evidences from simulation studies show that this class could also be a good proposal for reducing mean squared errors.

\textit{Keywords}: Informative prior, Prior-data conflict, Data-dependent prior, Mixture prior, Small sample size, Hierarchical approximation, Mean squared error.
\end{abstract}

\section{Introduction}

Prior elicitation is the core of every Bayesian analysis and the prior should represent the belief of the statistician before observing the data. But for several reasons in the last decades many attempts for including data information in the elicitation process have been proposed. Roughly speaking, the resulting data-dependent prior is just a prior that depends on the data and suffers from two main criticisms: data are used twice and the calculus of the Bayes' theorem may not be performed directly. 

Despite this evident contravention of the Bayesian philosophy, many statisticians dealt with the double use of the data in Bayesian inference, and many others use data-dependent priors for complex models. However, as invoked by \cite{wasserman2000asymptotic} almost twenty years ago, a theoretical justification for these distributions is missing and the need for data-dependent priors may become more common as the complexity for applied problems increases. Apparently, the call for the data-dependent Bayesians did not remain silent in these last years. As far as we can tell from reviewing the literature, we may recognize at least three frameworks for justifying the data-dependent approach within the Bayesian inference: the approximation of a hierarchical model through the estimation of some hyperparameters \citep{gelmandatadependent}; the definition of an adjusted data-dependent paradigm allowing for the Bayes' Theorem computation \citep{darnieder2011bayesian}; and the definition of a data-dependent prior as a measurable function from the data space $\mathcal{Y}^{m}$ to the set of priors $\mathcal{P}$ \citep{wasserman2000asymptotic}. In this paper we propose a class of data-dependent prior distributions that may be theoretically justified under all these frameworks. Moreover, the methodology presented in this paper turns out to be interpreted also in terms of a penalized likelihood framework \citep{cole2013maximum} for regression models, where the penalty term is the kernel of a prior distribution and the weight of such penalization is not fixed in advance ---as it happens for instance through cross-validation or empirical Bayes techniques.

Why proposing a new data-dependent prior formulation? We acknowledge at least two reasons. From a Bayesian point of view, we want to investigate the information's extent of a prior distribution, and our proposal follows the words of \cite{gelmanpriors}, when he says that we need a compromise between the information carried by a ``\textit{wildly unrealistic in most settings prior informative distribution and a noninformative prior, feasible only in settings where data happen to be strongly informative about all parameters}''. And from a broader statistical point of view, we are interested in the global quality of the model and on the assumptions we propose, and we believe our prior might be a good solution in case of model/prior misspecification.

According to the first argument, we are aware that the use of informative priors ---or, at least, weakly informative priors \citep{gelman2008weakly}--- is strongly encouraged by subjectivist Bayesians, especially when a prior information for a specific application is actually available. However, even if the model is simple, when the sample size is `small' it is not trivial to elicit an informative prior that does not dominate the inference. Using an informative prior distribution elicited from historical data ---as it is usual in medical studies, for instance--- could result in a mismatch between the prior and the observed data, the so called \textit{prior-data conflict} \citep{evans2006checking, mutsvari2016addressing}. Thus, it emerges clearly that measuring the information contained in a prior distribution is not referred only as a mathematical exercise, but turns out to be  helpful in terms of inference and prediction purposes. For instance, \cite{morita2008determining} developed the so called prior effective 
sample size (ESS), an index which measures the amount of 
information contained in a proposed prior distribution $\pi$ for 
the parameter $\theta$, computed with respect to a posterior 
$q_{m}(\theta|y)$  resulting from a baseline prior $\pi_{b}$, with $\pi_{b}$ less informative than $\pi$. When fitting a Bayesian model to a dataset consisting of 10 observations, an effective sample size of 1 is reasonable, whereas a value of 20 implies that the prior, rather than the data, dominates the inference: with a few data, there is the risk of being `too much informative'.


Motivated by these considerations, our method uses data for dealing directly with the priors construction. Given a pair of distributions consisting of an informative and a diffuse prior, our procedure measures the distance between the data at 
hand and an additional set of data generated under the informative prior until the resulting posteriors may be considered approximately equal. The corresponding value of such a distance ---bounded in the interval $[0,1]$--- is plugged into a two-components mixture of the prior distributions considered above. The greater is this value, the farther are the data (simulated and real) from the informative prior, 
and consequently the stronger is the influence of the diffuse 
prior in our specification.
We prove that the so obtained class of mixture data-dependent priors ---hereafter MDD priors--- satisfies some nice properties. Among these, the distributions of this class always have a closed form in conjugate models and preserve the conjugacy. Under mild  conditions, they yield a lower effective sample size than that provided by the informative prior ---substantially they provide less information. Moreover, evidences from simulation studies in the supplementary material accompanying this paper show that they also yield lower mean squared errors in presence of both model or prior misspecification.

It is worth noting that the use of mixture priors ---possibly with one relative precise component and the other more vague--- is not a novelty in Bayesian statistic. They have been introduced for making the inference robust in terms of a Bayesian perspective \citep{berger1986robust}, and developed for assessing any prior-data conflict \citep{schmidli2014robust,mutsvari2016addressing}. A mixture specification turns out to be useful also in Bayesian variable selection: a `spike and slab' prior \citep{miller2002subset} with fixed hyperparameters is assigned to the regression coefficients in the stochastic search variable selection approach ---see \cite{o2009review} for an overview on variable selection methods. 

The paper is organized as follows. Section~\ref{sec:datadep} reviews the existing data-dependent approaches and presents in a few details the frameworks proposed by  \cite{darnieder2011bayesian} and \cite{gelmandatadependent}; moreover, this section puts also in evidence the connection between the double use of the data and the penalized likelihood methods under a Bayesian perspective. In Section~\ref{sec:mixture} we introduce the MDD density class and describe the resampling algorithms required for building these priors. After introducing the notion of effective sample size, in Section~\ref{sec:theor} we focus on some theoretical results for the MDD priors; still, in this section we put in evidence the distribution-constant behaviour of the Hellinger distance in some special cases, if used as a data statistic. The information of the proposed class of priors is discussed in two examples for non standard models in Section~\ref{sec:case}: an exponential model with a Jeffreys prior and a logistic regression for determining the greatest amount of tolerable dose in phase I trial. Section~\ref{sec:concl} concludes.

\section{Using data twice in Bayesian inference}
\label{sec:datadep}

The commonly used expression `using data twice' in some Bayesian procedures does not mean nothing really precise, actually. However, it is not of interest for us taking an overview on all those tools which make use of the data twice for checking the fit of the model ---posterior predictive checkings, posterior Bayes fators, etc.--- or reviewing the empirical Bayes methods \citep{carlin2000bayes}. In this section we focus on those priors' procedures which explicitly consider data in the elicitation process. 

As widely known, using data or the data mechanism process in the priors' elicitation is not properly Bayesian and suffers from two main criticisms: using data twice and not allowing for the direct computation of the Bayes' Theorem. However, some authors have attempted to circumvent these criticisms. In what follows, we take a brief overview on some existing data-dependent approaches. Firstly, we present the theoretical framework proposed by \cite{darnieder2011bayesian}, who formalized the so called Adjusted Data-dependent Bayesian paradigm, a new approach which introduces an adjustment in order to obtain a proper Bayesian inference starting from a data-dependent prior. Then, we present and formalize the considerations presented by \cite{gelmandatadependent}, who proposed to approximate a hierarchical model by using a data-dependent prior. We refer at \cite{wasserman2000asymptotic} for the formulation of data-dependent priors that yield proper posteriors for finite mixture-models.

 Finally, we draw a parallel between data-dependent priors and the penalized likelihood methods commonly used in Bayesian variable selection. Although this paper does not explicitly take in consideration regression models, it is of future interest for us to implement our procedure also for regression purposes, and we consider this subsection as a grounding motivation for future work.

\subsection{Darnieder's approach}
\label{sec:darnieder}

Let $\bm{y}$ denote the sample of the data at hand, $\bm{\theta}$ the vector of parameters and $T( \bm{y})$ a statistic computed on the data. Let $\pi(\bm{\theta}| T( \bm{y}))$ denote a data-dependent prior whose dependence through the data is expressed by the statistic $T( \bm{y})$.  \cite{darnieder2011bayesian} espresses the joint probability density of $(\bm{\theta}, \bm{y}, T( \bm{y}))$ as:

\begin{align*}
p( \bm{\theta}, \bm{y}, T( \bm{y}))= & p( T( \bm{y})| \bm{\theta}, \bm{y})p( \bm{\theta}| \bm{y}) m( \bm{y}) \\
= & f(\bm{y}|\bm{\theta}, T( \bm{y})   )\pi(\bm{\theta}| T( \bm{y}))m( T( \bm{y}))
\end{align*}

where $m(\bm{y})$ is the marginal (or integrated) likelihood. By isolating the posterior distribution on the left side, we obtain

\begin{equation}
p( \bm{\theta}| \bm{y})= \frac{f(\bm{y}|\bm{\theta}, T( \bm{y})   )\pi(\bm{\theta}| T( \bm{y}))m( T( \bm{y}))}{p( T( \bm{y})| \bm{\theta}, \bm{y})m( \bm{y})}
\end{equation}

Now, we observe that given $\bm{y}$, $T( \bm{y})| \bm{\theta}, \bm{y} $ is not random, and that the ratio $m( T( \bm{y}))/m( \bm{y})$ depends only on the observed data. Hence, we may write the above expression as

\begin{equation}
p( \bm{\theta}| \bm{y}) \propto f(\bm{y}|\bm{\theta}, T( \bm{y})   )\pi(\bm{\theta}| T( \bm{y})).
\label{eq:naive_posterior}
\end{equation}

As stated by \cite{darnieder2011bayesian}, the posterior in~\eqref{eq:naive_posterior} is obtained through a \textit{naive} approach. The equation is suggesting that using a data-dependent prior requires that also the likelihood of the model should be conditioned on the statistic $T( \bm{y})$. This formula is mathematically appealing, but the update of $\pi(\bm{\theta}| T( \bm{y}))$ is often not straightforward. Hence, after some simple algebra, the posterior may be expressed as

\begin{equation}
p( \bm{\theta}| \bm{y}) \propto \frac{f(\bm{y}|\bm{\theta}   )\pi(\bm{\theta}| T( \bm{y}))}{ g( T( \bm{y})| \bm{\theta}) }=f(\bm{y}|\bm{\theta}   ) \frac{\pi(\bm{\theta}| T( \bm{y}))}{g( T( \bm{y})| \bm{\theta})}
\label{eq:adjusted_posterior}
\end{equation}

where the ratio $\pi(\bm{\theta}| T( \bm{y})/g( T( \bm{y})| \bm{\theta})$ is the actual data-dependent prior, updated with the usual unconditioned likelihood $f(\bm{y}|\bm{\theta})$. \cite{darnieder2011bayesian} defines the posterior in \eqref{eq:adjusted_posterior} as an \textit{adjusted} posterior, obtained through an adjusted procedure. He also shows a relationship between a genuine Bayesian approach and the data-dependent Bayesian approach, putting in evidence the following identity:

\begin{equation}
1= \frac{p( \bm{\theta}| \bm{y})m(\bm{y})}{f(\bm{y}|\bm{\theta}   )\pi(\bm{\theta})}=  \frac{\pi(\bm{\theta}| T( \bm{y}))m( T( \bm{y}))}{g( T( \bm{y})| \bm{\theta}) \pi(\bm{\theta})}
\label{eq:bayesian_comparison}
\end{equation}

By dividing this expression by the genuine prior $\pi(\bm{\theta})$, we can state the following proportionality, the so called data-dependent Bayesian Principle:

\begin{equation}
\frac{p( \bm{\theta}| \bm{y})}{f(\bm{y}|\bm{\theta}   )}\propto \frac{\pi(\bm{\theta}| T( \bm{y}))}{g( T( \bm{y})| \bm{\theta})}
\label{eq:bayesian_principle}
\end{equation}

which formally coincides with \eqref{eq:adjusted_posterior}, but suggests something even stronger. In fact, this expression highlights that the principle is satisfied whether a genuine prior $\pi(\bm{\theta})$ exists or not. With the adjusted procedure we provide a posterior distribution which is directly implied by Bayes' Theorem, whatever is the choice for $\pi(\bm{\theta})$.

A natural question concerns the choice of the statistic $T(\bm{y})$. There are no particular guidelines for choosing $T(\bm{y})$, but \cite{darnieder2011bayesian} lists some theorems that are useful for this aim. For example, it is trivial to show that if $T(\bm{y})$ is sufficient for $\bm{y}$, then the data-dependent prior $\pi(\bm{\theta}| T( \bm{y}))$ coincides with the genuine posterior $p( \bm{\theta}| \bm{y})$. And the following theorem in case of a distribution-constant statistic $T(\bm{y})$ will be useful later.

\begin{thm}
Suppose $T(\bm{y})$ is distribution-constant for $\bm{\theta}$, then the naive expression~\eqref{eq:naive_posterior} and the adjusted expression~\eqref{eq:adjusted_posterior} coincide. Furthermore, the data-dependent prior $\pi(\bm{\theta}| T( \bm{y}))$ coincides with the genuine prior $\pi(\bm{\theta})$.
\label{eq:thm_1}
\end{thm}

For a quick proof see the Appendix. As suggested by \cite{darnieder2011bayesian}, it is hard to imagine a beneficial conditioning on a distribution-constant statistic, unless for those priors which depend only on the data sample size. However, in Section~\ref{sec:theor} we will use this result for showing that, within some particular cases, our data-dependent prior procedure only depends on the sample size of our dataset and yields some good properties in terms of global information, frequentist coverage and mean squared errors.

\subsection{Gelman's approach}
\label{sec:gelman}

\cite{gelmandatadependent} draws an appealing framework considering the data-dependent priors as an approximation of a hierarchical model. He moves from a concrete example of regression models with standardized predictors: rescaling a bunch of predictors based on the data and then putting informative priors on their coefficients means eliciting a prior that depends on the data. He doesn't go in depth with mathematical notation, but we consider challenging to formalize this setup.

As usual in hierarchical models \citep{gelman2014bayesian}, let $\bm{y}$ represent the data-vector, $\bm{\theta}$ denote the generic vector of parameters and $\bm{\phi}$ the vector of hyperparameters. The likelihood of the model is $p(\bm{y}|\bm{\theta})$. The joint prior distribution for $(\bm{\theta}, \bm{\phi}) $ is

$$p(\bm{\theta}, \bm{\phi})= p(\bm{\phi})p(\bm{\theta}|\bm{\phi}), $$

and the joint posterior distribution is 

\begin{equation}
 p(\bm{\theta}, \bm{\phi}|\bm{y}) \propto p(\bm{\theta}, \bm{\phi}) p(\bm{y}| \bm{\theta}, \bm{\phi})=p(\bm{y}| \bm{\theta})p(\bm{\theta}|\bm{\phi})p(\bm{\phi}),
 \label{eq:hierarchical_model}
 \end{equation}

with the further assumption that the hyperparameter $\bm{\phi}$ affects $\bm{y}$ only through $\bm{\theta}$. In a full Bayesian model, $\bm{\phi}$ is not known and is assigned a prior distribution $p(\bm{\phi})$; however, in some circumstances it may be possible to consider $\bm{\phi}$ as known, or estimate it. As in the Gelman's example, if this hyperparameter, say a \textit{population} parameter, is estimated from the data, then we denote this estimate with ${\bm{\phi}}(\bm{y})$ and the population distribution $p(\bm{\theta}|\phi)$ reduces to $p(\bm{\theta} | {\phi}(\bm{y}) )$, which actually is a data-dependent prior according to \cite{darnieder2011bayesian}.
If we replace $\bm{\phi}$ with an estimate, $\bm{\theta}$ still preserves the dependence from  ${\bm{\phi}}(y)$, but the joint posterior distribution in~\eqref{eq:hierarchical_model} reduces to the following approximate hierarchical joint posterior,

\begin{equation}
 p(\bm{\theta}, \bm{\phi}(\bm{y})|\bm{y}) \propto p(\bm{\theta}| \bm{\phi}(\bm{y}), \bm{y}) p( \bm{\phi}(\bm{y})| \bm{y}) \propto p(\bm{\theta}| \bm{\phi}(\bm{y}), \bm{y}),  
 \label{eq:approx_joint}
 \end{equation}

where $p(\bm{\theta}| \bm{\phi}(\bm{y}), \bm{y})$ may be interpreted as the marginal approximate posterior for $\bm{\theta}$ ---analogous to the pseudo-posterior distribution in empirical Bayes methods \citep{petrone2014empirical}, where $\bm{\phi}(\bm{y})$ is usually obtained through marginal maximum likelihood estimation. We may derive an explicit form for this quantity by applying the Bayes' Theorem and the assumption $p(\bm{y}| \bm{\theta}, \bm{\phi}(\bm{y}))=p(\bm{y}| \bm{\theta})$:

\begin{equation}
p(\bm{\theta}| \bm{\phi}(\bm{y}), \bm{y}) \propto p(\bm{y}| \bm{\theta}, \bm{\phi}(\bm{y}))p(\bm{\theta}, \bm{\phi}(\bm{y}) ) \propto p(\bm{y}| \bm{\theta})p(\bm{\theta}|\bm{\phi}(\bm{y})).
\label{eq:appr_hierarchical_model}
\end{equation}

The comparison between this latter expression and~\eqref{eq:hierarchical_model},~\eqref{eq:approx_joint} highlights the relationship existing between a full Bayesian  hierarchical model and an approximate hierarchical model, where $\bm{\phi}(\bm{y})$ naturally acts in place of $\bm{\phi}$ and Bayes' Theorem is guaranteed by the product between the usual likelihood and the data-dependent prior $p(\bm{\theta}|\bm{\phi}(\bm{y}))$. The framework above has the merit of interpreting a data-dependent prior as an approximation of a further level of hierarchy within hierarchical models, through the use of a data-statistic $\phi(\bm{y})$ as a plug-in estimate for the hyperparameter $\phi$; moreover, it proposes the definition of a pseudo-posterior $p(\bm{\theta}| \phi(\bm{y}), \bm{y})$. 

\subsection{Penalized likelihood}
\label{sec:penalized}

In the penalized likelihood approaches for regression models ---Lasso \citep{tibshirani1996regression}, Ridge regression, Bridge regression--- it is usual to penalize some coefficients by inducing a certain amount of shrinkage in order to (i) overcome problems in the
stability of parameter estimates due to a relatively flat  likelihood and (ii) reduce the global mean squared error. A penalized log-likelihood with quadratic penalization is 

\begin{equation}
\log L(\bm{\beta}, \bm{y}) -\frac{r}{2}(\bm{\beta}-\bm{g})^{2},
\label{eq:penalized}
\end{equation}

where $\bm{\beta}=(\beta_{1},...,\beta_{J)}$ is the vector of regression parameters, $\bm{g}=(g_{1},...,g_{J})$ is a vector of values which should be good guesses for the vector parameter $\bm{\beta}$, and $(\bm{\beta}-\bm{g})^{2}= \sum_{j=1}^{J}(\beta_{j}-g_{j})^{2}$ is the quadratic penalty. The formula above may be easily interpreted in terms of a Bayesian perspective. In fact, if $\beta_{j} \sim \mathcal{N}(g_{j}, 1/r)$, then~\eqref{eq:penalized} represents a log-likelihood penalized by the log-density of the prior distribution for $\beta_{j}$, where $r$ is the precision (the inverse of the prior variance) and is usually called the \textit{tuning} parameter. Thus, the quadratic log-likelihood penalization reduces to eliciting independent normal priors on the parameters with prior mean $g_{j}$ and prior variance $1/r$. The ordinary Lasso of Tibshirani can be interpreted as a Bayesian Lasso \citep{park2008bayesian}, i.e. as a Bayesian posterior mode estimate when regression parameters have Laplace independent priors. And more generally Bridge regression is a direct generalization for Lasso and Ridge regression, where the penalty is $(\bm{\beta}-\bm{g})^{q}$ for some $q \ge 0$ ($q=1$ corresponds to the ordinary Lasso, $q=2$ to the Ridge regression). Many approaches for estimating the tuning parameter $r$ have been proposed: cross-validation, general cross-validation, empirical Bayes methods through marginal maximum likelihood estimation. But only assigning a diffuse hyperprior is purely Bayesian. Using data for estimating the tuning parameter makes in fact the Bayesian penalized log-likelihood approach affected by the data process and, more precisely, the prior on $\beta$ affected by the data. In Section~\ref{sec:theor} we put in evidence that our methodology allows for a hierarchical approximation and may be also justified in terms of log-likelihood penalization. 
 
%
%

\section{Mixture Data-dependent priors}
\label{sec:mixture}

Let $ \bm{y}_{m}= (y_{1},...,y_{m})$ be a data vector from a given sampling distribution $f(\bm{y}_{m}|\theta)$, with $\theta \in 
\mathbb{R}$. Let $\pi_{b}(\theta) $ denote a diffuse prior 
distribution for $\theta$ ---hereafter called \textit{baseline} prior--- and suppose that, from a preliminary knowledge about the problem (for instance historical information), we are somehow able to assign 
a more informative prior distribution $\pi(\theta)$. 
When data consist of a relatively small number of observations, the choice between these two priors' options is 
not trivial, since the support and the shape of the posterior are sensitive to the choice of the prior distribution. 
Thus, the information contained in the prior could turn out to be dominant when the dataset is small. This is one of the reasons for combining our previous information about the problem with our data at hand ----precisely, with an augmented version of it, as will be clarified later--- and proposing a data-dependent approach for eliciting a particular class of mixture prior distributions. We may then introduce the mixture data-dependent (MDD) prior $\varphi(\theta)$ with mixture weight $\psi_{m^{*}}$

\begin{equation}
\varphi(\theta)= \psi_{m^{*}}\pi_{b}(\theta)+(1-\psi_{m^{*}})\pi(\theta),
\label{eq:mixture:prior}
\end{equation}

belonging to the corresponding MDD class $$\Phi= \{ \varphi : \varphi(\theta)= \psi_{m^{*}}\pi_{b}(\theta)+(1-\psi_{m^{*}})\pi(\theta) ,\ \theta \in \Theta,\ 1 \ge\psi_{m^{*}}\ge 0 \}.$$ The MDD prior~\eqref{eq:mixture:prior} may then be viewed as a compromise between an informative prior and a noninformative one, with weights $\psi_{m^{*}}, \ 1-\psi_{m^{*}}$ obtained through a data augmentation with global length $m^{*}$. Note that mixture priors designed for overcoming the prior-data conflict and for robustness purposes have been already proposed by \cite{mutsvari2016addressing} and \cite{schmidli2014robust}: however, the authors do not propose any procedure for computing/assigning the mixture weights, and this is a crucial point for us, as explained in the next section.

\subsection{The resampling algorithms for the mixture weigths}
\label{sec:alg}

\begin{figure}
\centering
\includegraphics[scale=0.5]{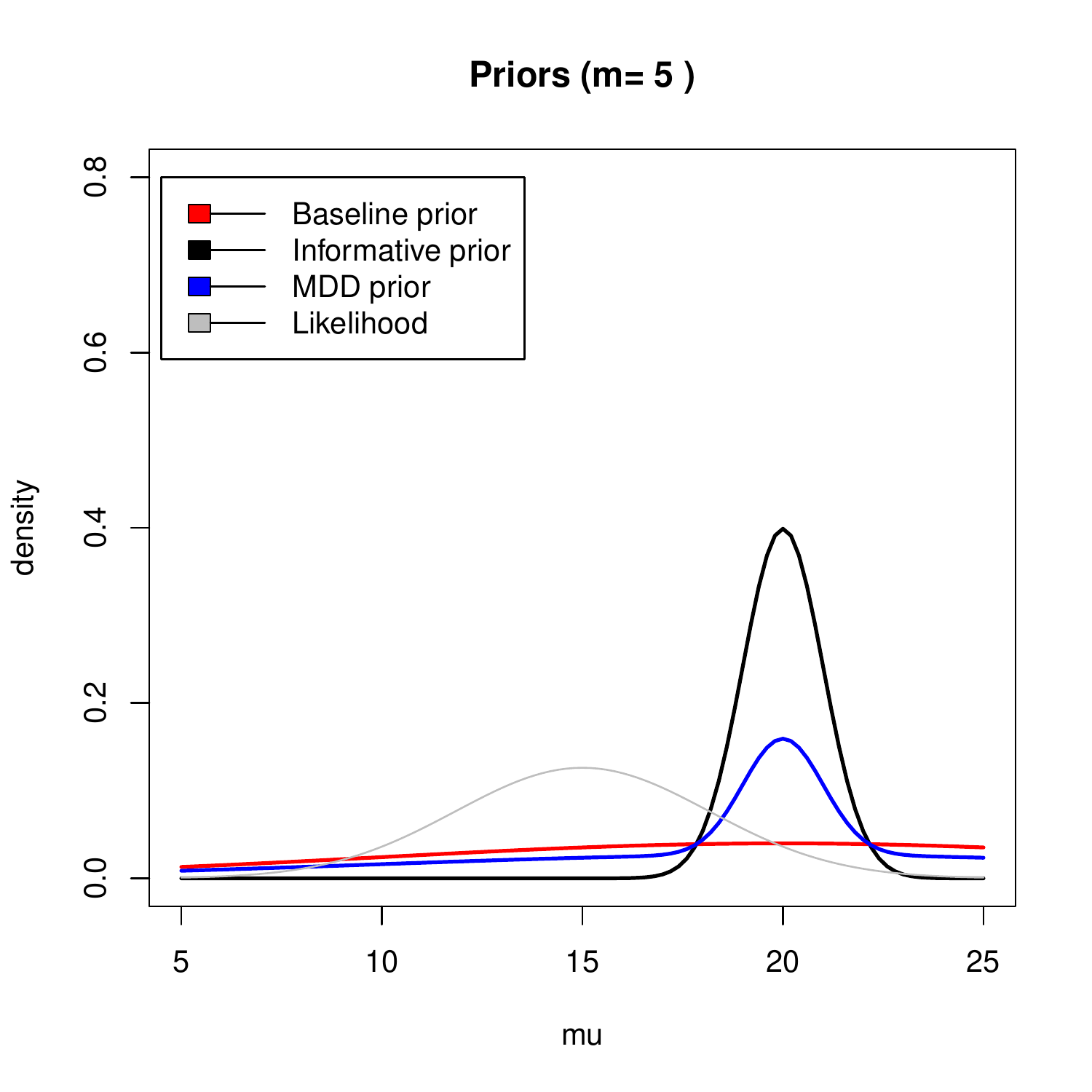}\\
\includegraphics[scale=0.5]{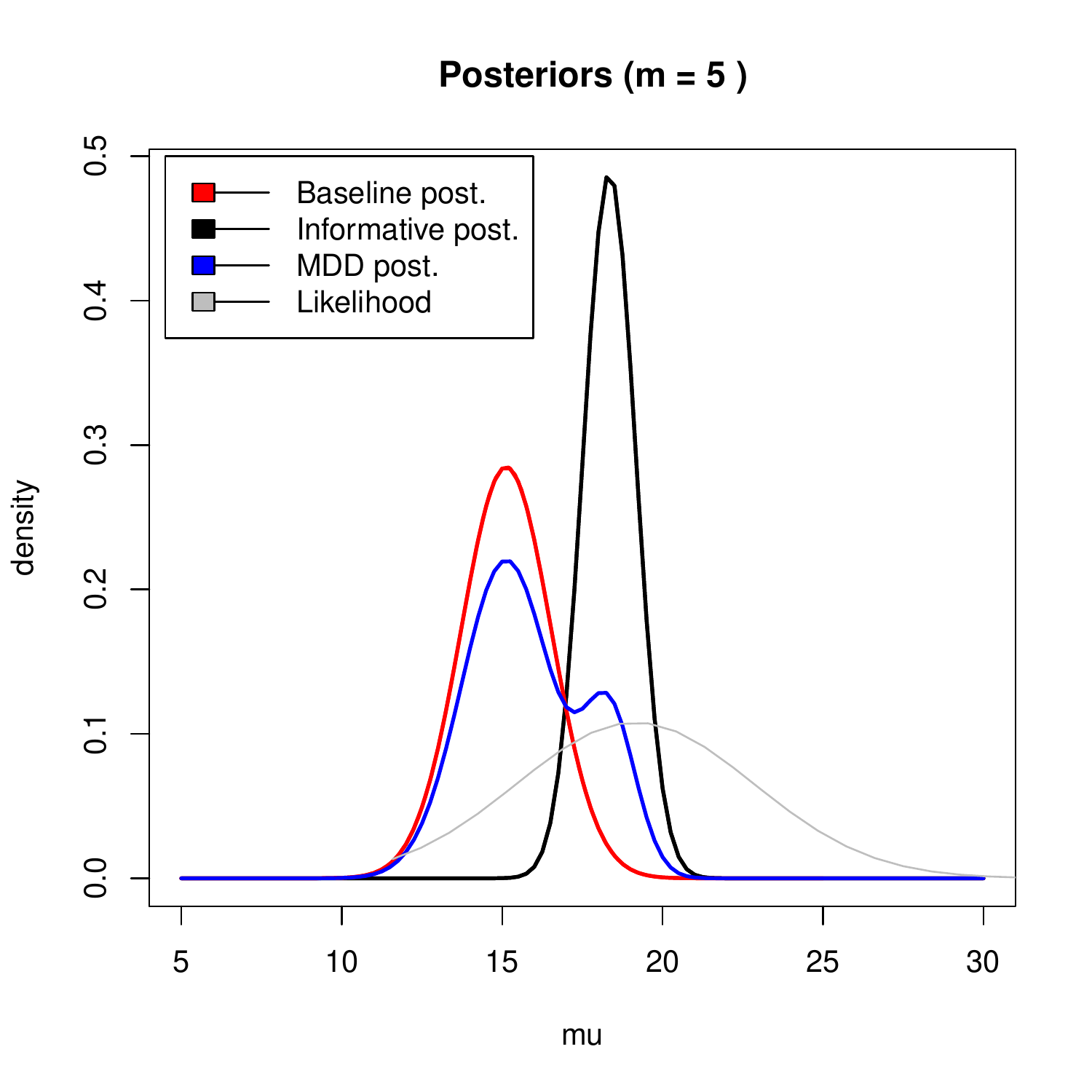}~
\includegraphics[scale=0.5]{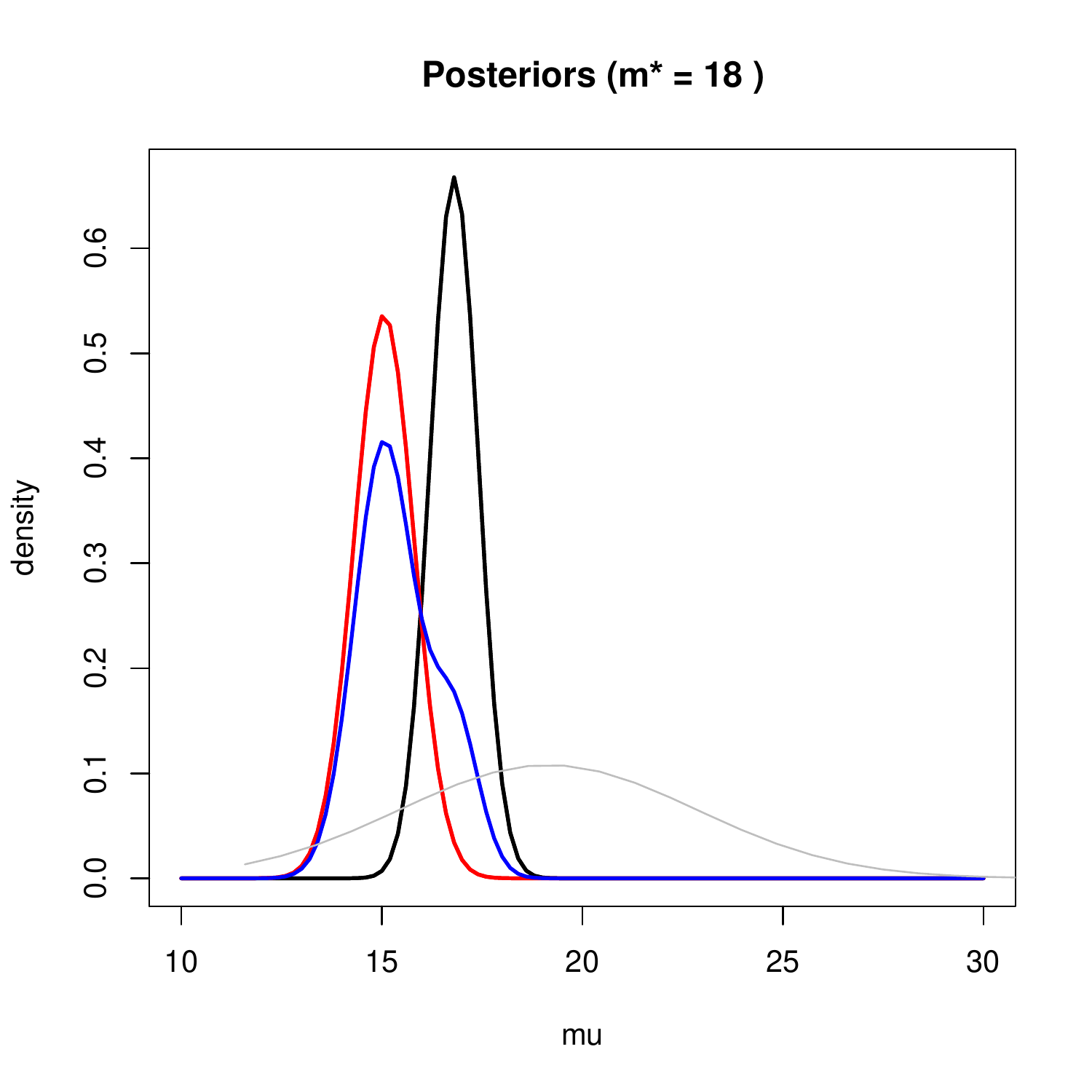}
\caption{Normal-Normal model, resampling-algorithm 1. (\textit{Top})  $f(\bm{y}_{m}|\theta)=\mathcal{N}(15,10)$ (grey line),  $\pi_{b}(\theta)=\mathcal{N}(20,100)$, $\pi(\theta)=\mathcal{N}(20,1)$ and $\varphi(\theta)=\psi_{m^{*}}\mathcal{N}(20,100)+(1-\psi_{m^{*}})\mathcal{N}(20,1)$. The initial sample is set to $m=5$. (\textit{Bottom row, left}) Baseline posterior $q_{m}(\theta|\bm{y}_{m})$, posterior $\pi_{m}(\theta|\bm{y}_{m})$, MDD posterior $\varphi_{m}(\theta|\bm{y}_{m})$ for the initial sample size $m $. The grey line is the density for the new values $\bm{y}_{\varkappa}$ generated under $f(\bm{y}_{m}| \theta^{*})$. (\textit{Bottom row, right}) Baseline posterior $q_{m^{*}}(\theta|\bm{y}_{m^{*}})$, posterior $\pi_{m^{*}}(\theta|\bm{y}_{m^{*}})$, MDD posterior $\varphi_{m^{*}}(\theta|\bm{y}_{m^{*}})$, for the sample size $m^{*}=m+\varkappa $, here 18.}
\label{normal_mixture}
\end{figure}

Assume to have observed the 
data vector $\bm{y}_{m}$, which represents our data at hand. Let simulate $\theta^{*} \sim \pi
(\theta)$ and define a modified version of the sampling distribution $f$ as $f(\bm{y}_{m}|\theta^{*})$. Assuming that $\theta_{0}$ is the true value of the parameter $\theta$ which generates our data at hand,  we compute the Hellinger distance $\mathcal{H}$ ---closely related to the Bhattacharyya distance   \citep{bhattacharyya1946measure}---  between our data generating process $f(\bm{y}_{m}| \theta_{0})$ and $f(\bm{y}_{m}|\theta^{*})$, defined as:

\begin{equation}
\Psi_{m} \equiv \mathcal{H}( f(\bm{y}_{m}|\theta_{0}), f(\bm{y}_{m}|\theta^{*}))=\frac{1}{\sqrt{2}} \left[ \int |\sqrt{f}-\sqrt{f^{*}}|^{2}d\bm{y_{m}} \right]^{\frac{1}{2}}
\label{eq:Hellinger:dist}
\end{equation}

where  $f^{*}$ is an abbreviate notation for $f(\bm{y}_{m}|\theta^{*})$. For any couple of density functions $g, h$, the Hellinger distance satisfies the property $0 \le \mathcal{H}(g,h) \le 1$.  It is worth noting that in~\eqref{eq:Hellinger:dist} we are treating $\theta_{0}$ as known, but in most of the statistical applications it is unknown and we need to estimate it. Among the others, one possibility could be that of using the maximum likelihood (ML) estimate $\hat{\theta}_{0}$, obtained equating at zero the log-derivative of the sampling distribution. Let $\psi_{m}$ denote the observed value of the Hellinger distance~\eqref{eq:Hellinger:dist}, bounded between 0 and 1. In an analogous way, let $\omega_{m}$ be the observed value of the Hellinger distance 

\begin{equation}
\Omega_{m} \equiv \mathcal{H}(q_{m}(\theta|\bm{y}_{m}), \pi_{m}(\theta|\bm{y}_{m}))
\label{eq:Hellinger:dist:K}
\end{equation}

 between the baseline posterior $q_{m}(\theta|\bm{y}_{m})$ and the informative posterior $\pi_{m}(\theta|\bm{y}_{m})$. The key-point of our procedure is that of sequentially generating $\varkappa$ new values $\bm{y}_{\varkappa}=(y_{m+1},...,y_{m+\varkappa})$, and re-computing the distances~\eqref{eq:Hellinger:dist},~\eqref{eq:Hellinger:dist:K} for each new draw, until a certain condition of similarity between the posterior distributions $\pi$ and $q$ is satisfied. Precisely, the stop condition is expressed by

\begin{equation}
\varkappa={\mbox{inf}}\ \{k \in \mathbb{N}\ | \Omega_{m+k} < \epsilon, \ \epsilon>0 \}
\label{eq:pess}
\end{equation}

for a fixed tolerance $\epsilon$. Thus, the so obtained $\psi_{m^{*}}$ is the observed value of $\Psi_{m^{*}}$, in correspondence of the dimension $m^{*}=m+\varkappa$ of the augmented dataset. This posterior similarity may be seen as an approximate matching between the proposed posterior distributions.  
 Note that the idea of matching the posterior uncertainty carried by two different posteriors doesn't represent a novelty, and a procedure based on the average posterior uncertainty is proposed by \cite{reimherr2014being}. The use of Hellinger distance is appropriate for some nice theoretical properties, as will be clarified in Section~\ref{sec:theor}.  

As mentioned above, a crucial point is the generation of the additional data. Given the specific problem at hand, there is not a unique way for achieving this task. We propose two possible procedures, respectively named resampling-algorithm 1 and resampling-algorithm 2: for a deep illustration of these methods see the Appendix. For illustration purposes only, Figure~\ref{normal_mixture} displays a graphical example for the mixture prior and posterior (blue lines) obtained through resampling-algorithm 1 for a simple Normal-Normal model. However, in both the procedures as $\psi_{m^{*}}$ approximates 1 (maximal distance), the mixture prior~
\eqref{eq:mixture:prior} approximates the baseline prior 
distribution $\pi_{b}(\theta)$; conversely, as $\psi_{m^{*}}$ 
approximates 0 (minimal distance), the mixture prior approximates the informative prior 
$\pi(\theta)$. In this formulation, the data dependence is expressed
by the presence in~\eqref{eq:mixture:prior} of the observed 
Hellinger distance $\psi_{m^{*}}$ between the actual and the further 
data at the $m^{*}$-th iteration. However, one could simply use the 
current set of data without the need of generating additional data. 
In such a case, Equation~\eqref{eq:mixture:prior} will be the same, 
but the weight $\psi_{m^{*}}$ may be 
computed as the observed value of the Hellinger distance between the 
informative prior $\pi(\theta)$ and the informative posterior 
$\pi_{m}(\theta|y_{m})$. Along the rest of the paper, we will refer to this 
formulation as the \textit{natural} MDD prior. Whereas MDD prior-res1 and MDD prior-res2 will denote respectively the MDD priors obtained with the resampling-algorithm 1 and 2.

\section{Theoretical results}
\label{sec:theor}

In this section we present some theoretical results for the MDD class presented in Section~\ref{sec:mixture} within the univariate conjugate models. Precisely, we introduce here the notion of effective sample size proposed by \cite{morita2008determining}, showing that the information of the MDD prior is always lower than the information of any informative prior. Moreover, we frame the MDD prior class in the theoretical approaches of \cite{darnieder2011bayesian} and \cite{gelmandatadependent}, summarized in Section~\ref{sec:datadep}. According to the first reference, we review the notion of  distribution-constant statistics and we put in evidence that in some special cases ---e.g. the Normal-Normal model, but generally all the statistical models for which the Fisher information doesn't depend on the parameter--- the Hellinger distance is a distribution-constant statistic. This property implies that in these special models our proposed methodology substantially reduces to choosing a genuine prior.

\begin{table}
\caption{\label{tab:01} $\theta \in \mathbb{R}$, $c\ge 1$. Suppose $\bm{y}_{m}=(y_{1},...,y_{m}) \sim f(\bm{y}_{m}|\theta)$.
 Prior $\pi(\theta)$, baseline prior $\pi_{b}(\theta)$, MDD prior $\varphi(\theta)$, 
 likelihood $f(\bm{y}_{m}|\theta) $, baseline posterior $q_{m}(\theta| \bm{y}_{m}) $ 
and MDD posterior $\varphi_{m}(\theta| \bm{y}_{m}) $ for the univariate conjugate models: Normal-Normal (NN), Gamma-Poisson (GP), Gamma-Exponential (GExp) and Beta-Binomial (BB). Following \cite{gelman2014bayesian}, 
we denote $\mathcal{N}(\mu, \sigma^2)$, $\mathcal{G}\mbox{a}(\alpha, \beta)$, $\mathcal{B}\mbox{e}(\alpha, \beta)$, $\mathcal{B}\mbox{in}(n, \theta)$, $\mathcal{P}\mbox{ois}(\theta)$ and $\mathcal{E}\mbox{xp}(\theta)$ for the normal, gamma, beta, binomial, Poisson and exponential distributions. 
For the Normal-Normal model let $\bar{\mu}(\tau^{2})= (\frac{\mu}{\tau^{2}}+\frac{m}{\sigma^{2}}\bar{y})/ (\frac{1}{\tau^{2}}+\frac{m}{\sigma^{2} })$ denote the posterior mean in function of the prior variance $\tau^{2}$, and $\bar{\tau}^{2}(\tau^{2})=(\frac{1}{\tau^{2}}+\frac{m}{\sigma^{2} })^{-1}$ the posterior variance in function of the prior variance $\tau^{2}$.}
\begin{small}
\begin{tabular}{|lll|}
\multicolumn{3}{c}{}\\
\hline
   &  \textit{NN} &  \textit{GP}  \\
   \hline\\
   \small $\pi_{b}(\theta)$ & $ \mathcal{N}(\mu, c\tau^{2})$ & $\mathcal{G}\mbox{a}( \frac{\alpha}{c}, \frac{\beta}{c})$ \\
   \small $\pi(\theta) $ & $ \mathcal{N}(\mu, \tau^{2})$ &$ \mathcal{G}\mbox{a}(\alpha, \beta)$  \\
  \small $\varphi(\theta)$ & $ \psi_{m^{*}} \mathcal{N}(\mu, c\tau^{2})+$& $\psi_{m^{*}}\mathcal{G}\mbox{a}(\frac{\alpha}{c}, \frac{\beta}{c})+$ \\
   & $ (1-\psi_{m^{*}})\mathcal{N}(\mu, \tau^{2})$ & $(1-\psi_{m^{*}})\mathcal{G}\mbox{a}(\alpha, \beta)$ \\
  \small $f(\bm{y}_{m}|\theta)$ & $\mathcal{N}(\theta, \sigma^{2})$ & $\mathcal{P}\mbox{ois}(\theta)$\\
  \small $q_{m}(\theta| \bm{y}_{m})$ &$ \mathcal{N}( \bar{\mu}(c \tau^{2}),\bar{\tau}^{2}(c\tau^{2})  )$ & $\mathcal{G}\mbox{a}(\frac{\alpha}{c}+\sum y_{i}, \frac{\beta}{c}+m)$ \\
  \small $\varphi_{m}(\theta| \bm{y}_{m})$ & $\psi_{m^{*}} \mathcal{N}( \bar{\mu}(c\tau^{2}),\bar{\tau}^{2}(c\tau^{2})  )+$ & $\psi_{m^{*}}\mathcal{G}\mbox{a}(\frac{\alpha}{c}+\sum y_{i}, \frac{\beta}{c}+m)+$ \\
  & $ (1-\psi_{m^{*}})\mathcal{N}( \bar{\mu}(\tau^{2}),\bar{\tau}^{2}(\tau^{2})  )$ & $(1-\psi_{m^{*}})\mathcal{G}\mbox{a}(\alpha+\sum y_{i}, \beta+m)$ \\ 
  \hline
  \multicolumn{3}{c}{}\\
\hline
   &  \textit{GExp} & \textit{BB} \\
   \hline\\
   \small $\pi_{b}(\theta)$ &  $\mathcal{G}\mbox{a}(\frac{\alpha}{c}, \frac{\beta}{c})$ & $\mathcal{B}\mbox{e}(\frac{\alpha}{c}, \frac{\beta}{c})$\\
   \small $\pi(\theta) $ &  $\mathcal{G}\mbox{a}(\alpha, \beta)$ & $\mathcal{B}\mbox{e}(\alpha, \beta)$\\
  \small $\varphi(\theta)$ &   $ \psi_{m^{*}}\mathcal{G}\mbox{a}(\frac{\alpha}{c}, \frac{\beta}{c})+$ & $\psi_{m^{*}}\mathcal{B}\mbox{e}(\frac{\alpha}{c}, \frac{\beta}{c})+$\\
   &  $(1-\psi_{m^{*}})\mathcal{G}\mbox{a}(\alpha, \beta)$ & $(1-\psi_{m^{*}})\mathcal{B}\mbox{e}(\alpha, \beta)$\\
  \small $f(\bm{y}_{m}|\theta)$ &   $\mathcal{E}\mbox{xp}(\theta)$&  $\mathcal{B}\mbox{in}(m, \theta)$\\
  \small $q_{m}(\theta| \bm{y}_{m})$ & $\mathcal{G}\mbox{a}(\frac{\alpha}{c}+m, \frac{\beta}{c}+m \bar{y})$  &$\mathcal{B}\mbox{e}(\frac{\alpha}{c}+m \bar{y}, \frac{\beta}{c}+m-m \bar{y})$\\
  \small $\varphi_{m}(\theta| \bm{y}_{m})$ &  $\psi_{m^{*}}\mathcal{G}\mbox{a}(\frac{\alpha}{c}+m, \frac{\beta}{c}+m \bar{y})+$ & $\psi_{m^{*}}\mathcal{B}\mbox{e}(\frac{\alpha}{c}+m \bar{y}, \frac{\beta}{c}+(m-m \bar{y}) )+$\\
  & $(1-\psi_{m^{*}})\mathcal{G}\mbox{a}(\alpha+m, \beta+m \bar{y})$ & $(1-\psi_{m^{*}})\mathcal{B}\mbox{e}(\alpha+m \bar{y}, \beta+m-m \bar{y})$\\ 
  \hline
   \end{tabular} 
   \end{small} 
   \end{table}

Before proceeding, we introduce here a general vector notation that turns out to be helpful in the following sections. Without loss of generality, let $\bm{\theta}$, $\bm{\theta} \in \mathbb{R}^{d}$, denote the parameters' vector, with $d \ge1$. Let the symbols  $\pi_{b}(\bm{\theta})$, $\pi(\bm{\theta})$  denote as before respectively  a baseline prior and an informative prior for $\bm{\theta}$. Let $m$ denote the generic sample size and $f( \bm{y}_{m}|\bm{\theta})$ the likelihood for our sample $\bm{y}_{m}=(y_{1},...,y_{m})$. Finally, let $q_{m}(\bm{\theta}|\bm{y}_{m})$ denote the baseline posterior for our parameter $\bm{\theta}$. In Section~\ref{sec:mixture} we used the symbols $m$ for the initial sample size, $\varkappa$ for the sample size of the generated sample of data and, consequently, $m^{*}=m+\varkappa$ for the global dimension of the data vector, comprising both the data at hand and those generated via resampling-algorithm 1 or 2. The MDD prior presented in this section obviously relies on $\psi_{m^{*}}$ and on a preliminary generation of $\varkappa$ values with one of the resampling algorithms introduced in~\ref{sec:alg}. The further technical assumptions are

\begin{align}
\begin{split}
 &E_{\pi_{b}}(\bm{\theta})=E_{\pi}(\bm{\theta})\\ 
& \mbox{Corr}_{\pi}(\theta_{i}, \theta_{j})= \mbox{Corr}_{\pi_{b}}(\theta_{i}, \theta_{j}), \  i \ne j \\
 & \mbox{Var}_{\pi_{b}}(\theta_{j})  >> \mbox{Var}_{\pi}(\theta_{j}), \ j=1,...,d.
  \end{split}
 \label{eq:assumption}
 \end{align}

\begin{table}
\caption{\label{tab:02} \small $\theta \in \mathbb{R}, \ c\ge 1, \ m$ is the generic sample size. Negative second derivatives of the log densities and 
effective sample sizes for the baseline prior $\pi_{b}(\theta)$, the informative prior $\pi(\theta)$ and the MDD prior 
 $\varphi(\theta)$, for the univariate conjugate models. Let $\bar{\theta}=E_{\pi}(\theta)$ denote the plug-in estimate. See Table~\ref{tab:01} for the priors' specification.}
\begin{small}
\begin{tabular}{|lllll|}
\hline
 & \textit{NN} &\textit{GP} &\textit{GExp}& \textit{BB} \\ 
\hline
 $D_{\pi_{b}}(\theta)$  & $1/c \tau^{2}$ & $\frac{(\alpha/c-1)}{\bar{\theta}^{2}}$ & $\frac{(\alpha/c-1)}{\bar{\theta}^{2}}$ & $(\frac{\alpha}{c}-1)\frac{1}{\bar{\theta}^{2}}+(\frac{\beta}{c}-1)\frac{1}{(1-\bar{\theta})^{2}}$ \\
 $D_{\pi} (\theta)$  & $1/ \tau^{2}$ & $(\alpha-1)\bar{\theta}^{-2}$ & $(\alpha-1)\bar{\theta}^{-2}$& $\frac{(\alpha-1)}{\bar{\theta}^{2}}+\frac{(\beta-1)}{(1-\bar{\theta})^{2}}$   \\
$D_{q}(m,\theta, \bm{y}_{m})$  &$m/\sigma^{2}$  & $\frac{(\alpha/c+\sum y_{i}-1)}{\bar{\theta}^{2}}$   & $\frac{(\alpha/c+m-1)}{\bar{\theta}^{2}}$ & $\frac{(\frac{\alpha}{c}+\sum_{i}y_{i}-1)}{\bar{\theta}^{2}}+ \frac{(\frac{\beta}{c}+m-\sum_{i}y_{i}-1)}{(1-\bar{\theta})^{2}}$   \\
$ESS(\pi_{b}(\theta))$ & $\sigma^{2}/c\tau^{2}$ & 0 & 0 & 0\\
$ESS(\pi(\theta))$ & $\sigma^{2}/\tau^{2}$ & $\frac{\alpha-\alpha/c}{\bar{y}}$ & $\alpha-\alpha/c$  & $\alpha+\beta$ \\
\hline
\end{tabular}
\end{small}
\end{table}

\subsection{Effective sample size (ESS)}
\label{sec:ess}

The idea of measuring and quantifying the amount of information contained in a prior distribution  is of a great theoretical appeal. Nevertheless, it has been not yet studied by many authors and many technical difficulties arise, including the impossibility of encompassing in a unique philosophical and mathematical framework the task of assessing the impact of a prior distribution: several distance measures and many definitions of prior sample size may be in fact adopted. In what follows we will refer to the work of \cite{morita2008determining}, who defined the prior effective sample size (ESS) of $\pi(\bm{\theta})$, with respect to the likelihood $f(\bm{y}_{m}|\bm{\theta})$ as that integer $m$ which minimizes the distance between $\pi(\bm{\theta})$ and the baseline posterior $q_{m}(\bm{\theta}|\bm{y}_{m})$. To define this distance, they used the second derivatives of the log densities (the observed informations)

\begin{equation}
D_{\pi,j}(\bm{\theta})= -\frac{\partial^{2} \log(\pi(\bm{\theta}))}{\partial \theta^{2}_{j}}, \ \ D_{q,j}(m, \bm{\theta}, \bm{y}_{m})=-\frac{\partial^{2} \log(q_{m}(\bm{\theta}|\bm{y}_{m}))}{\partial\theta^{2}_{j}}, \ j=1,...,d.
\label{eq:morita:derivative}
\end{equation}

In what follows, we will sometimes use the simplified notations $\pi, q_{m}$ in place of $\pi(\bm{\theta}), q_{m}(\bm{\theta}|\bm{y}_{m})$ and $D_{\pi,j},  D_{q_{m},j}$ in place of $D_{\pi,j}(\bm{\theta}), D_{q,j}(m, \bm{\theta}, \bm{y}_{m})$. 
Let $D_{\pi,+}=\sum_{j=1}^{d} D_{\pi,j}$ and $D_{q_{m},+}=\sum_{j=1}^{d} \int D_{q_{m},j}f(\bm{y}_{m})d\bm{y}_{m}$ denote the global information for the prior $\pi$ and the posterior $q_{m}$, respectively. The distance between the prior and the posterior for the sample size $m$ is then defined as

\begin{equation}
\delta(m, \bar{\bm{\theta}}, \pi, q_{m})=| D_{\pi,+}(\bar{\bm{\theta}})-D_{q_{m},+}(\bar{\bm{\theta}})|,
\label{eq:morita_distance}
\end{equation}

evaluated in $\bar{\bm{\theta}}=E_{\pi}(\bm{\theta})$, the prior informative mean. The ESS for $\pi$ is defined as

\begin{equation}
ESS(\pi(\bm{\theta}))=\underset{m \in \mathbb{N}}{\mbox{Argmin}} \{ \delta(m, \bar{\bm{\theta}}, \pi, q_{m})  \}.
\label{eq:morita:ess}
\end{equation}

When $d=1$, we will simply write $D_{\pi}, D_{q_{m}}$, suppressing the subscript `+'. Table~\ref{tab:01} shows an example of the priors and the posteriors for four univariate conjugate models: Normal-Normal, Gamma-Poisson, Gamma-Exponential and Beta-Binomial. Note that, under the assumptions in~\eqref{eq:assumption}, the baseline prior mean corresponds to the informative prior mean, and the hyperparameter $c$ is a large constant chosen to inflate the baseline variance.
 Table~\ref{tab:02} reports the  distances and the effective sample sizes for these univariate conjugate models. Similarly to the general expression in~\eqref{eq:morita_distance}, the distance between the MDD prior $\varphi(\theta)$ and the baseline posterior $q_{m}(\theta|\bm{y}_{m})$ evaluated in $\bar{\theta}=E_{\pi}(\theta)$ is defined as

\begin{equation}
\delta(m, \bar{\theta}, \varphi, q_{m})=| D_{\varphi}(\bar{\theta})-D_{q_{m}}(\bar{\theta})|,
\label{eq:egidi_distance}
\end{equation}

where $ D_{\varphi}$ has not in general a closed form and it is computed through an $\mathsf{R}$ routine. The effective sample size $ESS(\varphi(\theta))$ is computed for the MDD prior analogously as in~\eqref{eq:morita:ess}. For the univariate conjugate models the following theorem holds.

\begin{thm}
 Given $\theta \in \mathbb{R}$, the likelihood $f(\bm{y}_{m}|\theta)$, an informative prior $\pi(\theta)$, a baseline prior $\pi_{b}(\theta)$, the baseline posterior $q_{m}(\theta|\bm{y}_{m})$ and the MDD prior $\varphi(\theta)$ defined in~\eqref{eq:mixture:prior}, assume to be in a conjugate case and that the technical conditions in \eqref{eq:assumption}
  hold.
Then
\begin{equation}
ESS(\varphi(\theta)) \le ESS(\pi(\theta))
\label{eq:thesis}
\end{equation}
\label{eq:thm_2}
\end{thm}

Formula~\eqref{eq:thesis} provides an upper bound for the effective sample size of the MDD prior class, and yields an intuitive result. Although an analytic form of the ESS for this class of priors is not available, the interpretation is that whatever are the observed weights and the priors $\pi_{b}, \pi$ used in the formulation, the information contained in the MDD prior is never greater than the information contained in $\pi$. From a practical point of view, this prior distribution provides a lower information than that contained in the prior $\pi$, and is then more likely to not dominate the likelihood.

\subsection{Distribution-constant statistics}

In this section we frame the MDD priors approach within the general theoretical framework for the data-dependent priors proposed by \cite{darnieder2011bayesian} ---and summarized in Section~\ref{sec:datadep}--- and we draw an appealing theoretical comparison between the MDD priors and the Bayesian approach, under certain technical conditions.

 As alluded in Section~\ref{sec:datadep}, one of the key-points of the Darnieder's approach concerns the choice of the statistic $T(\bm{y})$ on which conditioning the prior distribution. As widely explained in Section~\ref{sec:mixture}, the MDD prior depends on the data only through the Hellinger distance defined in~\eqref{eq:Hellinger:dist}. For illustration purposes only and without loss of generality ---the theorems listed below preserve their validity in a multidimensional case--- let consider $\theta$ as a scalar parameter, $\theta \in \mathbb{R}$, and put $r(\theta, \theta + \triangle) \equiv \mathcal{H}( f(\bm{y}_{m}|\theta), f(\bm{y}_{m}|\theta+\triangle))$, where the parameters' difference $\triangle$  is not a parameter, but just an observed quantity which may be computed for each $m$, as $\triangle=\theta^{*}-\theta^{(0)}$ (see Section~\ref{sec:alg}). Let $I_{m}(\theta;f)=mI(\theta) $ denote the Fisher information for the parametric family $\{f(\bm{y}_{m}; \theta): \theta \in \Theta \}$ in case of independent observations. \cite{borovkovmathematical} state the following theorem.

\begin{thm}
If the function $\sqrt{f(\bm{y}_{m}|\theta)}$ is differentiable with respect to $\bm{\theta}$, and $I_{m}(\theta;f) $  is continuous, than there exists the limit:

\begin{equation}
\lim_{\triangle \rightarrow 0} \frac{r(\triangle)}{\triangle^{2}}=I_{m}(\theta;f) 
\end{equation}
\label{eq:thm_4}
\end{thm}


This Theorem provides a limiting behaviour for the Hellinger distance, as the difference $\triangle$ approximates zero. Furthermore, he also provides some uniform bounds for  $r(\triangle)/\triangle^{2}$:

\begin{thm}

If the parameters set $\Theta$ is compact, $f(\bm{y}_{m}| \theta) \ne f(\bm{y}_{m}| \theta+\triangle)$ whenever $\triangle >0$ and if $0 < I(\theta)\le h < \infty$ for a given constant $h$, then there exists a constants $g>0$ such that the following relation holds:

\begin{equation}
g < \frac{r(\triangle)}{\triangle^{2}}< h
\label{eq:unoform_bounds}
\end{equation}
\label{eq:thm_5}
\end{thm}

Theorem~\eqref{eq:thm_5} is stating that, for every choice of $\theta$, $ r(\triangle)$ is bounded between $g \triangle^{2}$ and $h \triangle^{2}$. Hence, denoting with $ \{ \theta^{(m)} \}$ a generic parameter sequence depending on the sample size $m$ and with $r^{(m)}(\triangle)$ the corresponding Hellinger distance, we may state the following corollary:

\begin{corl}

As $m \rightarrow \infty$, the distribution of $r^{(m)}(\triangle)$ doesn't depend on the parameter $\theta$ but only on the parameters' difference $\triangle$. 

\end{corl}

In our framework, the dependence on the data for the MDD class is expressed by the observed Hellinger distance $\psi_{m^{*}}$; thus,  we naturally set $T(\bm{y}_{m^{*}})= r^{(m^{*})}(\triangle)$. If $I_{m^{*}}(\theta;f)$ doesn't depend on the parameter $\theta^{(m^{*})}$ ---this happens for instance for the Normal, LogNormal, Cauchy and Logistic distributions--- then, as $m^{*} \rightarrow \infty$, the distribution of $T(\bm{y}_{m^{*}})$ doesn't depend on $\theta$, but only on the parameters' difference $\triangle$: in other words, $T(\bm{y}_{m^{*}})$ is distribution-constant and Theorem~\ref{eq:thm_1} in Section~\ref{sec:darnieder} holds. We may summarize these results and state the following theorem.

\begin{thm}

Given a parametric family of continuous distributions $\{f(\bm{y}_{m^{*}}| \theta), \theta \in \Theta \}$, if the Fisher information $I_{m^{*}}(\theta;f)$ doesn't depend on $\theta$, then the Hellinger distance $r^{(m^{*})}(\triangle)$ doesn't depend on $\theta^{(m^{*})}$ but only on the difference $\triangle$. This means that the statistic $T(\bm{y}_{m^{*}})= r^{(m^{*})}(\triangle) $ is distribution-constant and the MDD prior \eqref{eq:mixture:prior} $\pi(\theta | T(\bm{y}_{m^{*}}))$ reduces to the genuine prior $\pi(\theta)$.
\end{thm}

It is straightforward to show that, in this particular case, the MDD prior still depends on the data, but exhibits their dependence on the data only through conditioning on the sample size $m$, plus an augmented sample size $\varkappa$. And, as \cite{darnieder2011bayesian} suggests, there is no need of doing any adjustment, since the sample size $m$ is intrinsic in the likelihood and does not convey any information about $\theta$. However, preliminary simulation in the supplementary material show that conditioning on such a statistic yields some advantages in terms of frequentist coverage and mean squared errors, especially when the genuine prior distribution is not well posed. 

By concluding, we found some special cases that, due to the presence of distribution-constant statistics, may be reduced to a genuine Bayesian approach even conditioning the prior on a data statistic. 

\subsection{Approximation of a hierarchical model}

\begin{figure}
\centering
\includegraphics[scale=0.5]{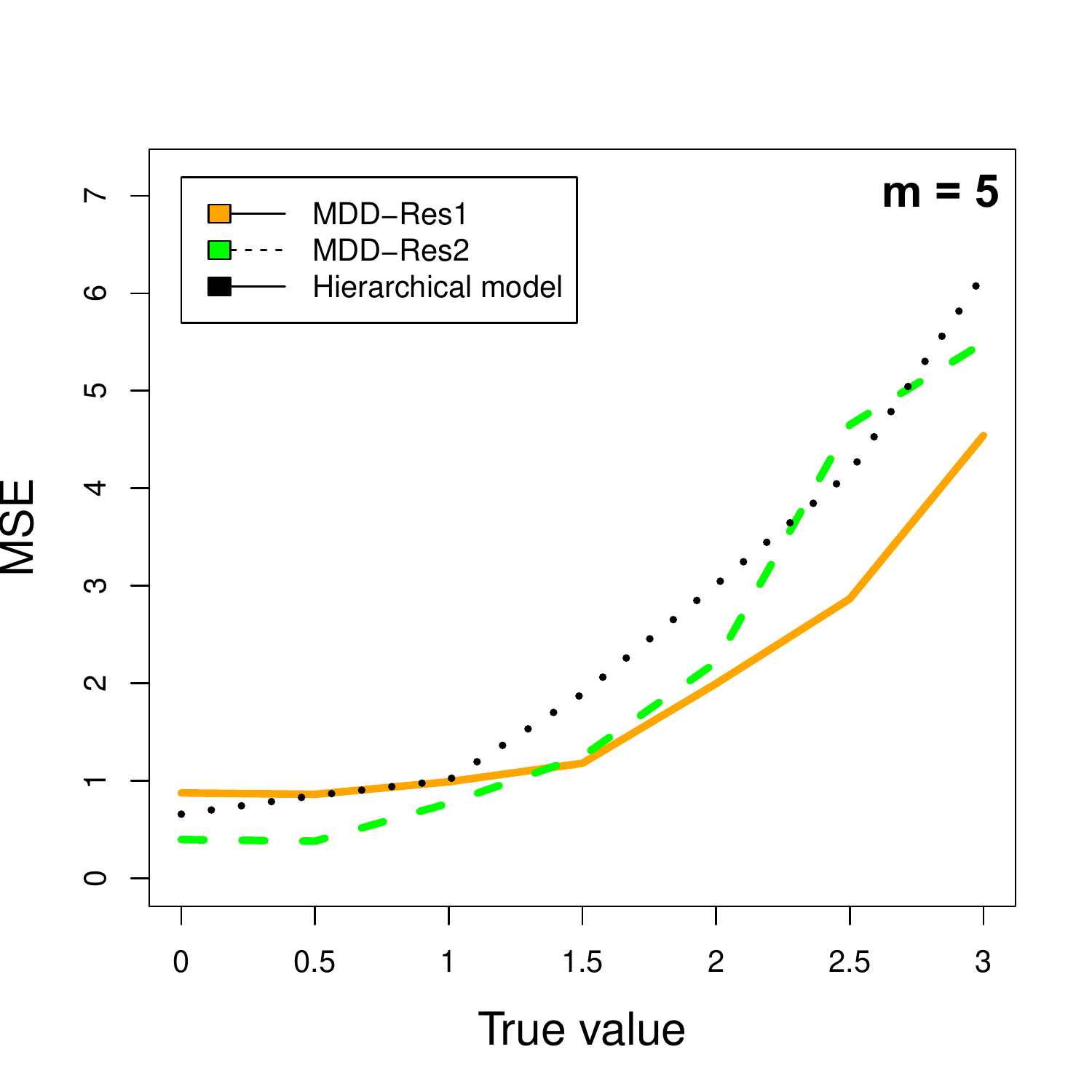}
\caption{Comparison between the MSE of the hierarchical model (dashed black line) and of the MDD prior-res1, MDD-prior-res2, with weights $\psi_{m^{*}}$ estimated from data. On the $x$-axis the true parameter value that generated the data. $c=100, \ \zeta^2=1,\  \sigma^2=5,\ m=5$. MSEs computed over 50 replications. The hierarchical model has been fitted using RStan \citep{rstan}, the R \citep{rcore} interface to
the Stan C++ library \citep{stan}.}
\label{hierarchical_comp}
\end{figure}

As suggested by \cite{gelmandatadependent}, data-dependent priors may sometimes be interpreted as an approximation of a hierarchical model, and in Sect.~\ref{sec:gelman} we provide a brief formalization of this intuition. Using again the Normal-Normal model as a toy example, let consider the following hierarchical model:

\begin{equation}
y_{ij}  \sim \mathcal{N}(\theta_{j[i]}, \sigma^{2}), \ i=1\ldots m, \ j=1,\ldots,J
\label{eq:hierarchical_model1}
\end{equation}
\begin{equation}
\theta_j \sim \mathcal{N}(0, \tau^{2}_{j})
\label{eq:hierarchical_model2}
\end{equation}
\begin{equation}
 \tau^{2}=\begin{cases}
\zeta^{2} \ \  \mbox{with } p \\
c\zeta^{2}  \ \mbox{with } 1-p
\end{cases}
\label{eq:hierarchical_model3}
\end{equation}

where the nested index $j[i]$ codes as usual in the hierarchical models \citep{gelman2006data} the group membership for the statistical unit $i$; the group-level parameter $\theta_{j}$ is assigned a normal prior distribution; the prior variance $\tau^2$ may assume two different values with probabilities $p$ and $1-p$; $c,\zeta^2$ are for simplicity fixed hyperparameters. If we fit this model according to the Bayesian paradigm, we should also assign a prior distribution to the probability $p$, for instance $p\sim \mathcal{B}\mbox{e}(a,b)$, depending on some hyperparameters $a,b$. The MDD prior for $\theta$, $\theta \sim p\mathcal{N}(0, c\zeta^2)+(1-p)\mathcal{N}(0, \zeta^2)$, is another way for expressing equations~\eqref{eq:hierarchical_model2},~\eqref{eq:hierarchical_model3}. We may then argue that the MDD class is a natural approximation of the model above, with the parameter $p$ that is not assigned a prior but estimated from the data through the resampling algorithms in Section~\ref{sec:alg}. 
For illustration purposes only, Figure~\ref{hierarchical_comp} displays a comparison, obtained through simulation using RStan \citep{rstan}, the R \citep{rcore} interface to
the Stan C++ library \citep{stan}, between the mean squared errors obtained from the hierarchical model in~\eqref{eq:hierarchical_model1},~\eqref{eq:hierarchical_model2},~\eqref{eq:hierarchical_model3}, the MDD prior-res1, and the MDD prior-res2, with $c=100, \ \zeta^2=1,\ \sigma^2=5,\ m=5$. The MDD priors show lower MSEs as the true value $\theta_{0}$ moves away from zero, the prior mean.

\subsection{Model for the tuning parameter}

As mentioned in Sect.~\ref{sec:penalized}, the relationship of the penalized likelihood to Bayesian theory is explained by the penalty through the kernel of the prior log-density. However, the estimation of the penalty weight remains open. \cite{hastie2002elements} suggest to use cross-validation, whereas \cite{efron2012large} propose empirical Bayes methods. Otherwise, \cite{cole2013maximum} set different values and examine the results for these different inputs. The MDD prior specification may be seen as a natural alternative for estimating the tuning parameter in the penalized likelihood approach. For illustration purposes only, let consider the regression model

$$y_{i}=\beta_{0}+\sum_{j=1}^{J}\beta_{j}x_{ij} +\epsilon_{i},$$

where $\epsilon_{i}\sim \mathcal{N}(0, \sigma^{2})$. And consider now the penalized log-likelihood with quadratic penalty for this model

\begin{equation}
\l(\bm{\beta}; \bm{y}) -\frac{1}{2\tau^{2}}\bm{\beta}^{2},
\label{eq:penalized_hierarchical}
\end{equation}

where $\beta_{j} \sim \mathcal{N}(0, \tau^{2})$ according to the Bayesian interpretation of the Ridge regression. The penalty weight/tuning parameter is $r=1/\tau^{2}$, the inverse of the prior variance. Instead of estimating directly this factor, specifying a MDD prior for $\beta_{j}$ is an automatic tool for introducing an auxiliary level for the variance, as in~\eqref{eq:hierarchical_model3}, and estimating the proportion $p$ through the resampling algorithms in Sect.~\ref{sec:alg}:

\begin{align}
\l(\bm{\beta}; \bm{y})& -\frac{1}{2\tau^{2}}\bm{\beta}^{2}\\
 \tau^{2}=&\begin{cases}
\zeta^{2} \ \  \mbox{with } p \\
c\zeta^{2}  \ \mbox{with } 1-p.
\end{cases}
\label{eq:penalized_hierarchical2}
\end{align}

Although we use the Normal-Normal model, this approach allows flexibility also for other types of prior distributions \citep{wood2017generalized}.

The penalized methods ---Lasso, Ridge regression, etc.--- are designed for reducing the mean squared errors, and the MDD class of priors, together with the resampling algorithms, represents a built-in method for addressing the same objective. Further work should be developed in order to implement the MDD priors for regression models and within the Bayesian variable selection framework.

\section{Examples to Some Nonstandard Models}
\label{sec:case}

In the previous sections we dealt with a pair of priors $\pi$ and $\pi_{b}$ belonging to the same family of distributions, under the technical condition in~\eqref{eq:assumption}. This is the same choice adopted by \cite{morita2008determining} and allows for inflating the noninformative variance by a factor $c$ and falling into the conjugate models. However, one may be interested in exploring other prior choices for $\pi_{b}$, possibly automatic priors, and attempting to measure the information carried by the MDD prior~\eqref{eq:mixture:prior}, by taking unchanged the informative prior $\pi$. In this section we explore this possibility and we focus on the corresponding amount of priors' information through a toy example and through a real case from a phase I trial study.

\subsection{Jeffreys prior for an exponential model}
\label{sec:Jeffreys_exp}

Let $\bm{y}_{m}=(y_{1},..., y_{m}) \underset{iid}{\sim} \mathcal{E}xp(\theta)$, with $\pi(\theta)= \mathcal{G}a(\alpha,\beta)$. The likelihood is then

\begin{equation}
L_{m}(\theta; \bm{y}_{m})=\prod_{i=1}^{m} f(y_{i})= \theta^{m} \exp(-\theta \sum_{i} y_{i}).
\label{eq:exponential_lik}
\end{equation} 

We introduce the Fisher information for the exponential model computed for a single observation:

$$I_{\theta}=E \left[- \frac{d^{2} \log f(y; \theta)}{d \theta^{2}} \right]=$$
$$= -E \left[ \frac{d^{2}}{d \theta^{2}} \left[  \log(\theta)-\theta y\right] \right]=-E \left[\frac{d}{d \theta}[1/{\theta}-y ] \right]=E \left[ \frac{1}{\theta^{2}} \right]= \frac{1}{\theta^{2}}.$$

Let $\pi_{b}(\theta)=j(\theta)$, where $j(\theta)= I^{1/2}_{\theta}$ is the Jeffreys prior. 
For the exponential model, the Jeffreys prior for $\theta$ is

\begin{equation}
j(\theta)=I^{1/2}_{\theta}=1/\theta.
\label{eq:Jeffreys_exp}
\end{equation}

Now we compute the Jeffreys posterior $q_{m}(\theta|y_{1},...,y_{m})= j_{m}(\theta|y_{1},...,y_{m} )$:

\begin{equation}
j_{m}( \theta| \bm{y}_{m}) \propto j(\theta) L_{m}(\theta; \bm{y}_{m})
=\theta^{-1}\prod_{i=1}^{m}\theta\exp \{-\theta y_{i} \}
=\theta^{m-1} \exp \{ -\theta \sum_{i=1}^{m} y_{i}  \}
\label{eq:Jeffreys_post}
\end{equation}

We immediately realize that this is the kernel of a Gamma distribution, $\mathcal{G}a(m, \sum_{i} y_{i})$

$$ j_{m}(  \theta |\bm{y}_{m} )= \frac{(\sum_{i}y_{i})^{m}}{\Gamma(m)} \theta^{m-1} \exp \{ -\theta \sum_{i=1}^{m} y_{i}  \}. $$

 We compute the negative second log derivative of $j_{m}(\theta |\bm{y}_{m})$ and we find the familiar result for a Gamma distribution

\begin{equation}
D_{j_{m}}=-\frac{d^{2}}{d \theta^{2}}\left[j_{m}( \theta| \bm{y}_{m})  \right]= \frac{m-1}{\theta^{2}}
\label{eq:Morita_Jeffreys}
\end{equation}

\begin{figure}
\centering
\subfloat[$\psi=0.2$]
{\includegraphics[scale=0.5]{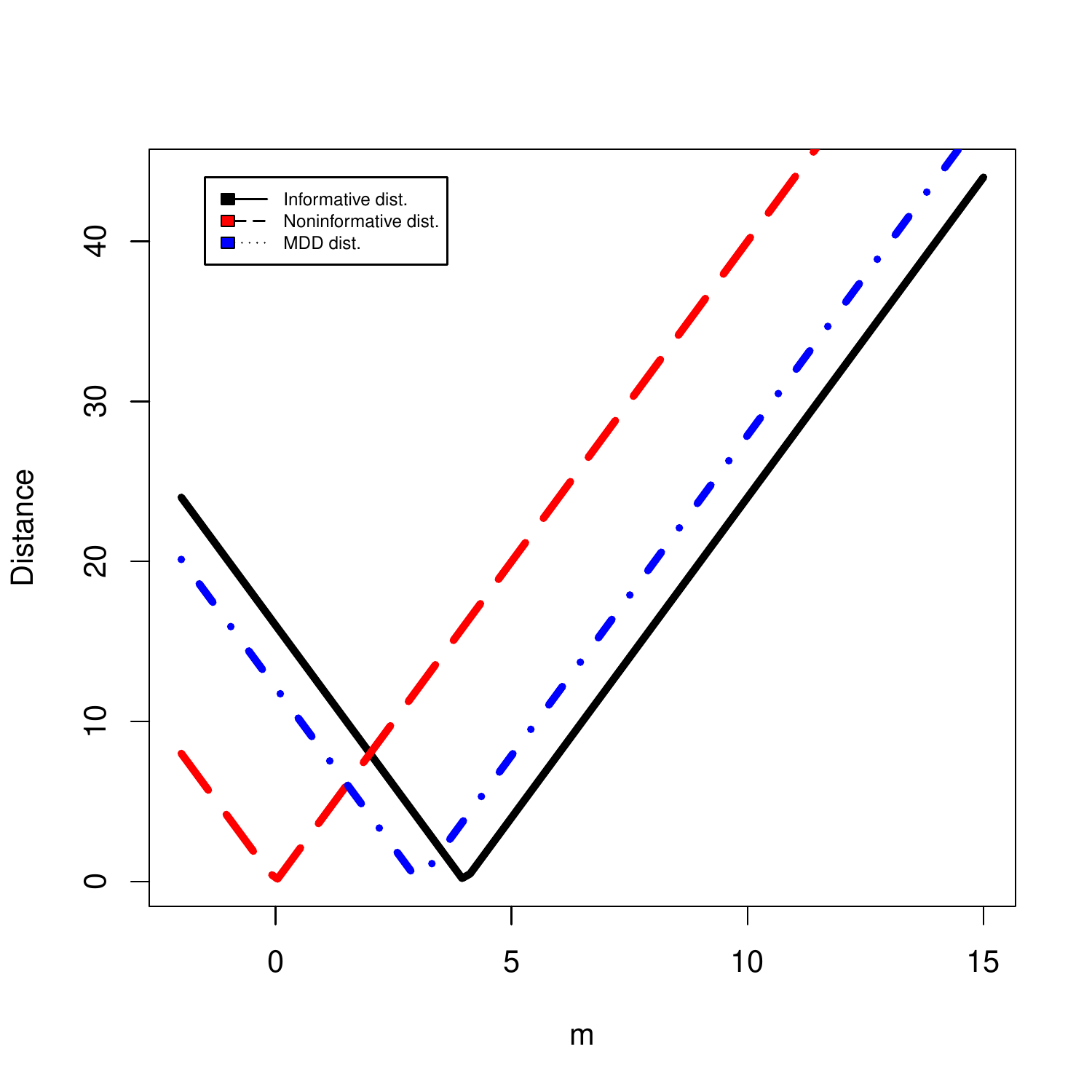}}~
\subfloat[$\psi=0.5$]
{\includegraphics[scale=0.5]{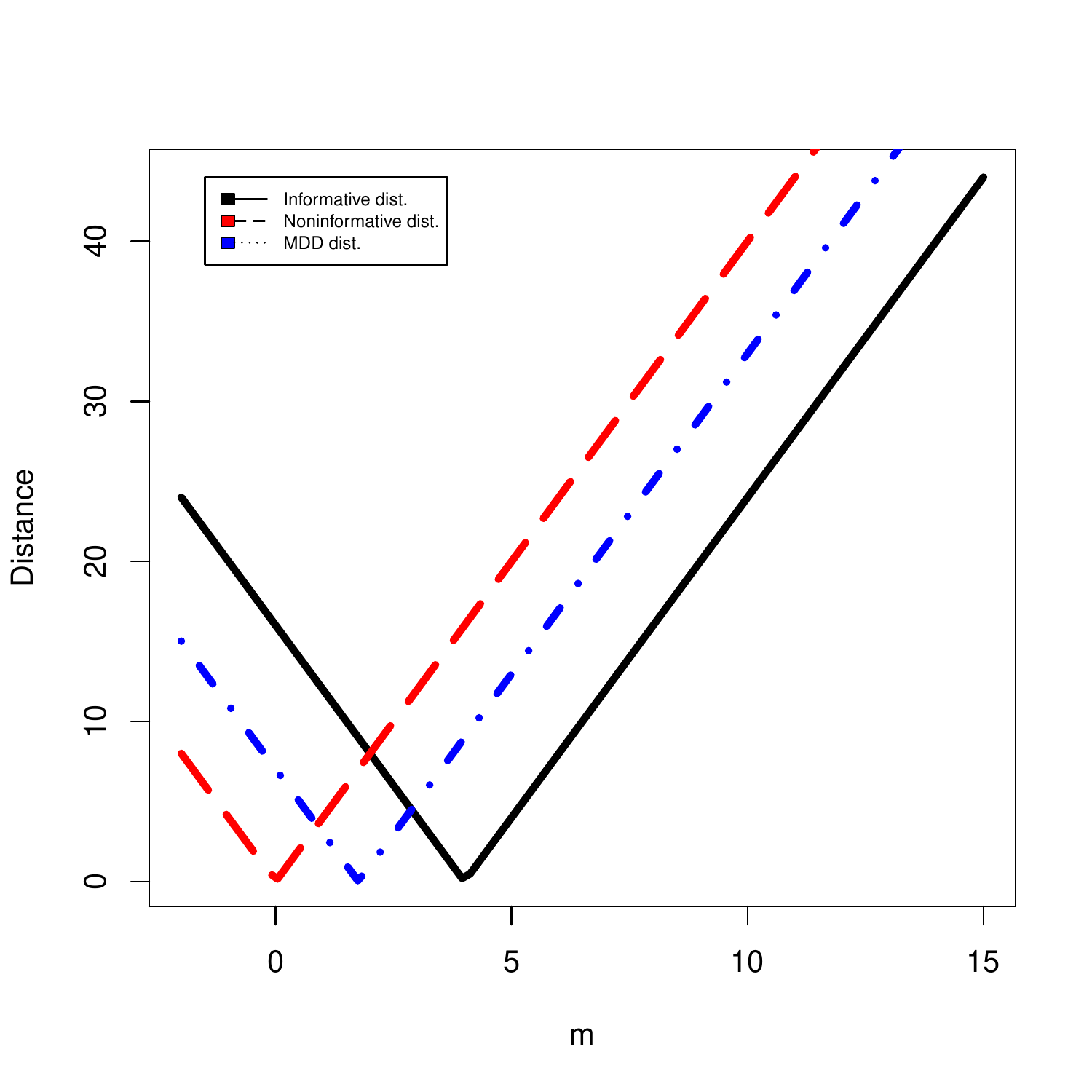}}\\
\subfloat[$\psi=0.8$]
{\includegraphics[scale=0.5]{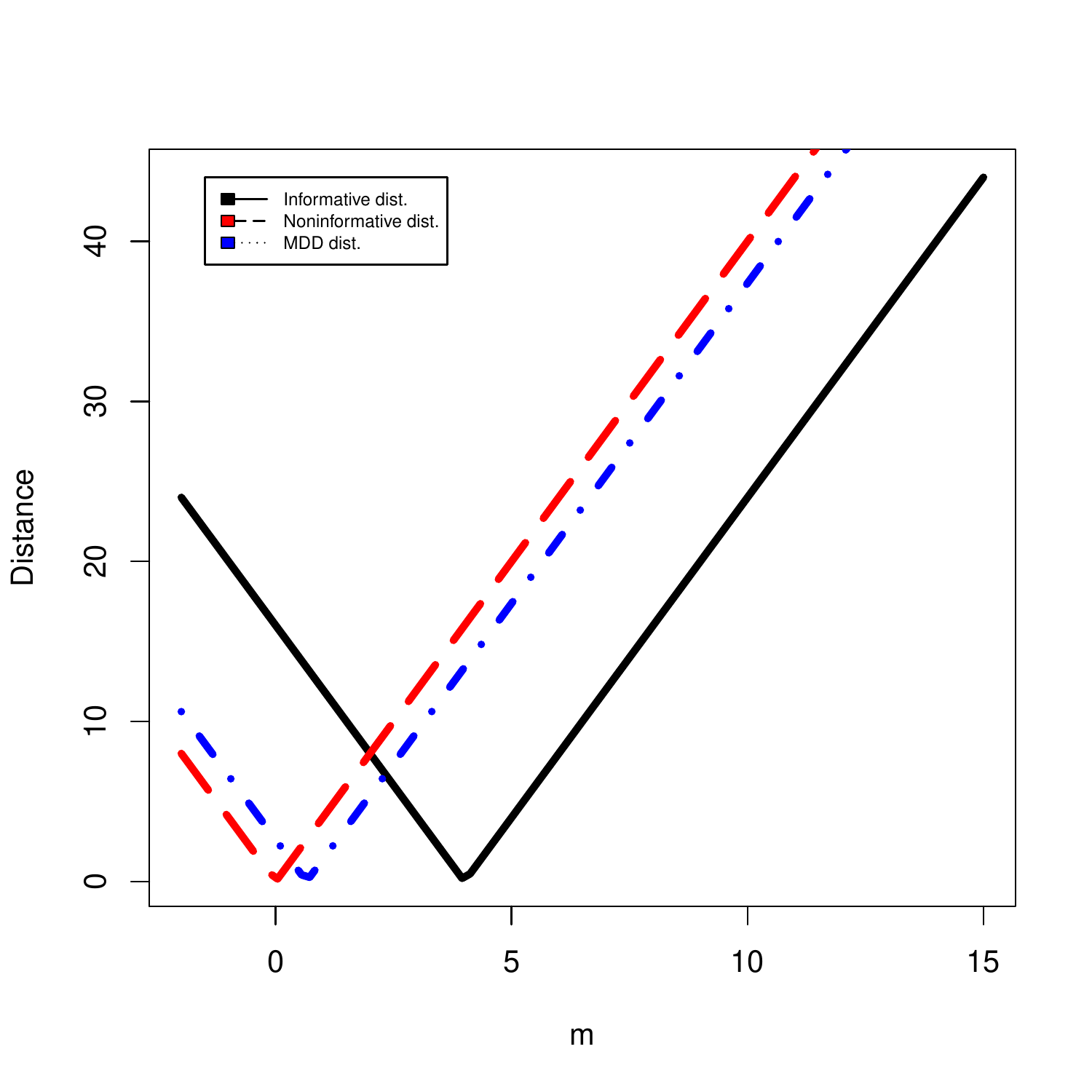}}
\caption{Exponential model with Jeffreys prior: on $y-$axis the distances $\delta(m, \bar{\theta}, \pi, q_{m})$ for the prior $\pi$ (solid black line), $\delta(m, \bar{\theta}, \pi_{b}, q_{m})$ for the baseline prior $\pi_{b}$ (dashed red line) and $\delta(m, \bar{\theta}, \varphi, q_{m})$ for the mixture prior $\varphi$ (dashed blue line) plotted against the sample size on $x-$axis. All these distances are minimized in correspondence of their effective sample size.}
\label{Jeffreys_esp}
\end{figure}

Finally, by using the plug-in estimate $\bar{\theta}=\alpha/\beta$, we may compute: 1) the distance~\eqref{eq:morita_distance} between the informative prior $\pi$ and the Jeffreys posterior $j_{m}$; 2) the distance between the Jeffreys prior $j$ and the Jeffreys posterior $j_{m}$; 3) the distance~\eqref{eq:egidi_distance} between the MDD prior $\varphi$ and the Jeffreys posterior $j_{m}$. Fig.~\ref{Jeffreys_esp} shows these distances according to three different values for the Hellinger distance, where the informative prior is set to $\pi(\theta)= \mathcal{G}a(4,8)$. The distance for $\varphi$ is always bounded between the distances of $j$ and $\pi$: hence, the ESS ---the value which minimizes these quantities--- for $\varphi$ is bounded between the effective sample sizes respectively for $j$ and $\pi$. As is intuitive, as the mixture weight increases, $ESS(\varphi(\theta))$ approximates $ESS(j(\theta))$. 

\subsection{Logistic regression for phase I trial}

\cite{thall2003practical} proposed a logistic regression to determine the greatest amount of tolerable dose in a phase I trial. In this section we follow the approach of \cite{morita2008determining}, who used the same example for studying the properties of the effective sample size for different values of the hyperparameters. 

The level of dose which each patient may receive is one among 100, 200, 300, 400, 500, 600 mg/m$^{2}$, denoted by $x_{1},\ldots,x_{6}$. These values are  then standardized on the log scale and denoted with $X_{1},...,X_{6}$. The response variable is $y_{i}=1$ if patient $i$ suffers toxicity, $y_{i}=0$ if not. They assume the following logistic model:

\begin{equation}
P(y_{i}=1) \equiv \pi(X_{i}, \bm{\theta})=logit^{-1}(\mu+\beta X_{i}) , \ i=1,...,m
\label{eq:logit}
\end{equation}

where $logit^{-1}(x)=e^{x}/(1+e^{x})$. Unlike the conjugate models considered in Section~\ref{sec:ess}, here the dimension of the parameters' space is $d=2$, $\bm{\theta}=(\mu, \beta)$, where $\mu$ is the intercept of the linear predictor and $\beta$ is the coefficient associated to the different levels of the doses. In order to compute the effective sample size, we need the extension to the multivariate case outlined by \cite{morita2008determining}. The likelihood for a sample of $m$ patients $\bm{y}_{m}=(y_{1},...,y_{m})$ is

\begin{equation}
f(\bm{y}_{m}| X, \bm{\theta})=\prod_{i=1}^{m}\pi(X_{i}, \theta)^{y_{i}}(1-\pi(X_{i}, \theta))^{1-y_{i}}
\label{eq:likelihood_logistic}
\end{equation}

\cite{thall2003practical} elicited two independent informative priors for $\mu$ and $\beta$ based on preliminary sensitivity analysis:

\begin{align}
\begin{split}
\mu &\sim \pi(\mu)=\mathcal{N}(\tilde{\mu}_{\mu}, \tilde{\sigma}^{2}_{\mu})=  \mathcal{N}(-0.11313, 2^{2})\\
  \beta & \sim \pi(\beta)= \mathcal{N}(\tilde{\mu}_{\beta}, \tilde{\sigma}^{2}_{\beta})= \mathcal{N}(2.3980, 2^{2}).
  \end{split}
  \label{eq:phaseI_thall_priors}
  \end{align} 
  
Hence, the baseline posterior is $q_{m}(\bm{\theta}| \bm{y}) = \mathcal{N}(\tilde{\mu}_{\mu}, c\tilde{\sigma}^{2}_{\mu})\mathcal{N}(\tilde{\mu}_{\beta}, c\tilde{\sigma}^{2}_{\beta})$, where the hyperparameter $c$ is fixed at 10000. We follow the steps of the algorithm formulated by \cite{morita2008determining} for determining (i) the effective sample size of each subvector and (ii) the global effective sample size of the parameter vector $\bm{\theta}$ as those values which respectively minimize the distances $\delta_{1}(m_{\mu},\bar{\bm{\theta}}, \pi_{\mu}, q_{m_{\mu}}), \delta_{2}(m_{\beta},\bar{\bm{\theta}}, \pi_{\beta}, q_{m_{\beta}})$ and $ \delta(m,\bar{\bm{\theta}}, \pi, q_{m})$, by using the plug-in vector $\bar{\bm{\theta}}=(\tilde{\mu}_{\mu}, \tilde{\mu}_{\beta})$. See the Appendix for a deep illustration of the algorithm. In this way, we compute the effective sample size of each parameter's subvector and then the global effective sample size of the logistic model. Given the two priors  $\pi_{\mu}, \pi_{\beta}$ in~\eqref{eq:phaseI_thall_priors}, we will denote the first two quantities with $ ESS(\pi(\mu)), ESS(\pi(\beta))$, and the third one simply with $ESS$. Table~\ref{tab:03} in the Appendix reports these effective sample sizes, obtained replicating the experiment of \cite{morita2008determining} and evaluated with respect to different values of the priors variances $\sigma^{2}_{\mu}, \sigma^{2}_{\beta}$. As intuitive, the information contained in the prior distributions decreases as the variances increase. In any case, the parameter $\beta$, associated to 
the effect of the doses, yields a greater knowledge than the parameter $\mu$, which represents the average response.
We repeat the same steps above adopting our mixture data-dependent prior by specifying for the vector parameter $\bm{\theta}$ the priors 

\begin{align}
\begin{split}
\mu \sim & \ \varphi(\mu)= \psi \mathcal{N}(\tilde{\mu}_{\mu}, c\tilde{\sigma}^{2}_{\mu})+ (1-\psi)\mathcal{N}(\tilde{\mu}_{\mu}, \tilde{\sigma}^{2}_{\mu})\\
\beta \sim & \ \varphi(\beta)= \psi \mathcal{N}(\tilde{\beta}_{\beta}, c\tilde{\sigma}^{2}_{\beta})+ (1-\psi)\mathcal{N}(\tilde{\mu}_{\beta}, \tilde{\sigma}^{2}_{\beta})
\end{split}
\label{eq:phaseI_priors}
\end{align}

 where the hyperparameter $c$ is fixed at 10000 as before and $\psi$ is the mixture weight. Being in absence of actual data at hand, here we do not adopt the algorithms of Section~\ref{sec:alg} for computing the observed value $\psi_{m^{*}}$ of the Hellinger distance $\Psi_{m^{*}}$: thus, for illustration purposes only, we drop the subscript $m^{*}$ and we consider three different values for $\psi$, $\psi= \{0.2,0.5,0.8 \}$. Then, we compare the so obtained results with those obtained with the above mentioned prior distributions.  As may be noticed from Table~\ref{tab:04}, as $\psi$ increases the effective sample sizes for the MDD priors~\eqref{eq:phaseI_priors} slightly decrease, as expected. However, the values obtained under these mixture priors are quite close to those obtained under the above priors $\pi(\mu),\  \pi(\beta)$ originally chosen by \cite{thall2003practical}. It would be worth assessing how much varies the information of the mixture priors $\varphi$ by choosing other baseline priors instead of flat normal distributions. Let us consider two improper priors, $\pi_{b}(\mu) \propto 1, \ \pi_{b}(\beta) \propto 1$. The resulting mixture priors $\varphi(\mu), \ \varphi(\beta)$ are then defined as
 
 \begin{align}
 \begin{split}
\mu \sim & \ \varphi(\mu)= \psi + (1-\psi)\mathcal{N}(\tilde{\mu}_{\mu}, \tilde{\sigma}^{2}_{\mu})\\
\beta \sim & \ \varphi(\beta)= \psi + (1-\psi)\mathcal{N}(\tilde{\mu}_{\beta}, \tilde{\sigma}^{2}_{\beta}).
\end{split}
\label{eq:phaseI_priors2}
\end{align}

Table~\ref{tab:05} in the Appendix reports the effective sample sizes for the priors in~\eqref{eq:phaseI_priors2}. In this case, there is an evident decrease of the information associated to the mixture priors $\varphi$: as $\psi$ increases and the improper priors are then preferred, the effective sample size rapidly decreases. This is intuitive, since the improper priors which appear in~\eqref{eq:phaseI_priors2} provide less information than two flat normal priors in~\eqref{eq:phaseI_priors}. The example suggests that even inflating the noninformative variances by a great factor $c$ doesn't affect in a sensible way the amount of information contained in the mixture prior. We may conclude that the best way for reducing an extra amount of information is combining an informative prior with an improper or ---when possible--- with a Jeffreys prior as in Section~\ref{sec:Jeffreys_exp}.

\section{Concluding remarks}
\label{sec:concl}

In this paper a new class of data-dependent prior distributions is proposed. This class consists of a two-component mixture of a baseline (flat) prior $\pi_{b}$ and an informative prior $\pi$, weighted through resampling methods in such a way to prefer $\pi_{b}$ if the additional set of data generated under $\pi$ appears to be far from the data at hand. This prior turns out to be a good proposal for avoiding prior-data conflict in presence of small sample size and first evidences from simulation studies suggest good performances for reducing the mean squared errors. 
 
Using the notion of effective sample size within conjugate models, we proved that the MDD prior class always provides a lower information than an informative prior.

Furthermore, different solutions for eliciting the baseline prior $\pi_{b}$ are explored: flat prior belonging to the same family of $\pi$, Jeffreys prior, improper prior. As is just partially intuitive, different strategies for the noninformative prior yield different extents of information for the MDD prior. 

Further work should be done in many directions. We should in fact explore more complex models, whose a brief sketch is only outlined in this paper.  Performing a proper sensitivity test for the selected priors $\pi_{b}, \pi$ is also a task of future interest. Finally, we strongly believe that extending the proposed methodology for regression models in terms of Bayesian variable selection is one crucial point in future research.

\bibliographystyle{chicago}
\bibliography{biblioarxiv}

\section*{Appendix}

\subsection*{Resampling algorithms}

\begin{small}

 According to the resampling-algorithm 1, we directly generate a sample $\bm{y}_{\varkappa}=(y_{m+1},...,y_{m^{*}}) $ from $f(\bm{y}_{m}| \theta^{*})$. At each step $k, \ k=1, \ldots, \varkappa$, we compute the Hellinger distances 

\begin{align}
\begin{split}
\Psi_{m+k} \equiv & \mathcal{H}( f(\bm{y}_{m}|\theta_{0}),  \bm{y} _{m+k}  )  \\
\Omega_{m+k} \equiv &\mathcal{H}(q_{m+k}(\theta|\bm{y}_{m+k}), \pi_{m+k}(\theta|\bm{y}_{m+k})),
\end{split}
\label{eq:Hellinger_m1}
\end{align} 
where the first equation in~\eqref{eq:Hellinger_m1} is the Hellinger distance between an absolute continuous distribution $f(\bm{y}_{m}|\theta_{0})$ and a numerical sample of length $m+k$ \footnote{We used the R function {\tt{HellingerDist}} of the {\tt{distrEx}} package \citep{kohl2007distrex}.}; whereas in the second equation $ q_{m+k}(\theta|\bm{y}_{m+k})$ denotes the baseline posterior computed in correspondence of the sample size $m+k$. 

According to the resampling-algorithm 2, we generate the values $\bm{y}_{\varkappa}=(y_{m+1},...,y_{m^{*}}) $ from the  sampling distribution $f(\bm{y}_{m}| \theta_{0})$ and at each step we compute the Hellinger distances

\begin{align}
\begin{split}
\Psi_{m+k} \equiv & \mathcal{H}( f(\bm{y}_{m}|\hat{\theta}^{(k)}_{0}), f(\bm{y}_{m}| \theta^{*}) )\\
\Omega_{m+k} \equiv &\mathcal{H}(q_{m+k}(\theta|\bm{y}_{m+k}), \pi_{m+k}(\theta|\bm{y}_{m+k})),
\end{split}
\label{eq:Hellinger_m2}
\end{align} 

where the Equation~\eqref{eq:Hellinger_m2} is the Hellinger distance between two absolute continuous distributions, and $\hat{\theta}^{(k)}_{0}$ the ML estimate for $\theta_{0}$ at step $k$, based on $y_{1},\ldots,y_{m},$ $y_{m+1},\ldots,y_{m+k}$.

 Resampling-algorithm 1 implies a data generation from the informative prior and compares these further data with those at hand: in some sense, this method is actually checking whether the informative prior is close to the data generating process. Perhaps, this procedure is oriented to assess the \textit{prior misspecification}. While in the resampling-algorithm 2, the data are generated according to the true model $f$:  this second algorithm assesses the \textit{model misspecification}.

\framebox[\textwidth][l]{\parbox{\textwidth}{\textbf{Resampling-algorithm 1}:
\\

Given $y_{1},...,y_{m} \sim f(\bm{y}_{m}| \theta)$, generate $\theta^{*} \sim \pi
(\theta)$. 

Fix the tolerance $\epsilon$.

Given $\Psi_{m} \equiv \mathcal{H}( f(\bm{y}_{m}|\theta_{0}), f(\bm{y}_{m}|\theta^{*}))$ and $\Omega_{m} \equiv \mathcal{H}(q_{m}(\theta|\bm{y}_{m}), \pi(\theta|\bm{y}_{m}))$ compute the observed\\ values $\psi_{m}$,  $\omega_m$. If the true value $\theta_{0}$ is unknown, provide an estimate for it.

Set $k=1$.

$\ \ \ \Diamond$  generate $y_{m+k}$ from $f(\bm{y}_{m}|\theta^{*})$. Given \begin{align*}
\Psi_{m+k} &\equiv \mathcal{H}( f(\bm{y}_{m}|\theta_{0}), \bm{y} _{m+k}  )\\
\Omega_{m+k} &\equiv \mathcal{H}(q_{m+k}(\theta|\bm{y}_{m+k}), \pi_{m+k}(\theta|\bm{y}_{m+k}))
\end{align*} 

$\ \ \ \Diamond\Diamond$ Compute the observed values $\psi_{m+k}, \omega_{m+k}$.

$\ \ \ $ \textbf{while} $\{\omega_{m+k} > \epsilon\}$ set  $k=k+1$ and go back to $\Diamond$.

Save $\psi_{m+\varkappa}$, $\omega_{m+\varkappa}$ and the new sample size $m^{*}=m+\varkappa$.
Set the prior~\eqref{eq:mixture:prior} with $\psi_{m^{*}}$.
}}

\framebox[\textwidth][l]{\parbox{\textwidth}{\textbf{Resampling-algorithm 2}:
\\

Given $y_{1},...,y_{m} \sim f(\bm{y}_{m}| \theta)$, generate $\theta^{*} \sim \pi
(\theta)$. 

Fix the tolerance $\epsilon$.

Given $\Psi_{m} \equiv \mathcal{H}( f(\bm{y}_{m}|\theta_{0}), f(\bm{y}_{m}|\theta^{*}))$, $\Omega_{m} \equiv \mathcal{H}(q_{m}(\theta|\bm{y}_{m}), \pi(\theta|\bm{y}_{m}))$ compute the observed values $\psi_{m}$,  $\omega_m$. If the true value $\theta_{0}$ is unknown, provide an estimate for it.

Set $k=1$.

$\ \ \ \triangle$  generate $y_{m+k}$ from $f(\bm{y}_{m}|\hat{\theta}^{(k)}_{0})$. Given \begin{align*}
\Psi_{m+k} &\equiv \mathcal{H}( f(\bm{y}_{m}|\hat{\theta}^{(k)}_{0}), f(\bm{y}_{m}| \theta^{*}) )\\
\Omega_{m+k} &\equiv \mathcal{H}(q_{m+k}(\theta|\bm{y}_{m+k}), \pi_{m+k}(\theta|\bm{y}_{m+k}))
\end{align*} 
$\ \ \ $ with $\hat{\theta}^{(k)}_{0}$ the ML estimate for $\theta_{0}$ at step $k$.

$\ \ \ \triangle\triangle$ Compute the observed values $\psi_{m+k}, \omega_{m+k}$.

$\ \ \ $ \textbf{while} $ \{\omega_{m+k} > \epsilon \}$ set  $k=k+1$ and go back to $\triangle$.

Save $\psi_{m+\varkappa}$, $\omega_{m+\varkappa}$ and the new sample size $m^{*}=m+\varkappa$.
Set the prior~\eqref{eq:mixture:prior} with $\psi_{m^{*}}$.
}}
   
\end{small}

\subsection*{Proof of Theorem \ref{eq:thm_1}}

Due to distribution-constant definition, $g( T( \bm{y})| \bm{\theta})= g( T( \bm{y}))$ and then $$p( \bm{\theta}| \bm{y}) \propto  f(\bm{y}|\bm{\theta}   )\pi(\bm{\theta}| T( \bm{y}))/ g( T( \bm{y})| \bm{\theta})  \propto f(\bm{y}|\bm{\theta}   )\pi(\bm{\theta}| T( \bm{y})).$$

Furthermore, $ \pi(\bm{\theta}| T( \bm{y})) \propto g( T( \bm{y})| \bm{\theta}) \pi(\bm{\theta}) \propto \pi(\bm{\theta}). \ \Box$

\subsection*{Proof of Theorem \ref{eq:thm_2}}
\begin{small}
\textit{Proof.}  For simplicity of notation we denote with $\alpha$ the baseline prior $\pi_{b}(\theta)$, with $\gamma$ the informative prior $\pi(\theta)$ and with $\beta$ the mixture prior $\varphi(\theta)= \psi_{m^{*}}\pi_{b}(\theta)+(1-\psi_{m^{*}})\pi(\theta)$. Furthermore, we abbreviate the weight $\psi_{m^{*}}$ as $\psi$. Unless otherwise stated, the dependence of the quantities introduced in Section \ref{sec:theor} on the parameter $\theta \in \mathbb{R}$ is here implicit. We compute the negative second log-derivative for the mixture prior \eqref{eq:mixture:prior} in general terms as 

\begin{align}
D_{\varphi}= &-\frac{d^{2} \log \{\varphi(\theta)\}}{d \theta^{2}}= -\frac{d^{2} \log \{\psi\pi_{b}(\theta)+(1-\psi)\pi(\theta) \}}{d \theta^{2}} =\\
& = -\frac{d}{ d\theta} \left[ \frac{\psi\alpha^{'}+(1-
\psi)\gamma^{'}}{\psi\alpha +(1-\psi)\gamma } \right]=\\
& =\frac{(\psi\alpha^{'}+(1-
\psi)\gamma^{'})^2- (\psi\alpha^{''}+(1-
\psi)\gamma^{''}) (\psi\alpha +(1-\psi)\gamma )}{(\psi\alpha +(1-\psi)\gamma )^2} 
\label{eq:D}
\end{align}

After some simple expansions we can rewrite \eqref{eq:D} and apply some minorations:

$$ D_{\varphi}=  \frac{ \psi^2 [ (\alpha^{'})^2- \alpha^{''}\alpha ] 
+(1- \psi)^2 (\gamma^{'})^2+2\psi(1- \psi) \gamma^{'}\alpha^{'}}{(\psi\alpha +(1-\psi)\gamma )^2} -$$
$$-\frac{\psi(1- \psi)\alpha^{''}\gamma+ \psi(1- \psi)\alpha\gamma^{''}+(1- \psi)^2 \gamma
\gamma^{''} }{(\psi\alpha +(1-\psi)\gamma )^2} \le $$
$$ \le \left[\frac{(\alpha^{'})^2- \alpha^{''}\alpha}{\alpha^2} \right]+ \frac{(1- \psi)^2 (\gamma^{'})^2-(1- \psi)^2 \gamma
\gamma^{''}}{(1-\psi)^2\gamma^2}+$$
$$+\frac{2\psi(1- \psi) \gamma^{'}\alpha^{'}-\psi(1- \psi)\alpha^{''}\gamma- \psi(1- \psi)\alpha\gamma^{''}}{\psi^2\alpha^2}=$$

\begin{equation}
=D_{\alpha}+ K_{1}
\label{diseq:D1}
\end{equation}

where $ K_{1}$ collects all the terms which do not enter in $D_{\alpha}$. Analogously, we can find another minoration:

$$ D_{\varphi} \le \left[\frac{(\gamma^{'})^2- \gamma^{''}\gamma}{\gamma^2} \right]+ \frac{\psi^2 (\alpha^{'})^2-\psi^2 \alpha
\alpha^{''}+2\psi(1- \psi) \gamma^{'}\alpha^{'}}{\psi^2\alpha^2}-$$
$$-\frac{\psi(1- \psi)\alpha^{''}\gamma+ \psi(1- \psi)\alpha\gamma^{''}}{\psi^2\alpha^2} =$$

\begin{equation}
=D_{\gamma+ K_{2}}
\label{diseq:D2}
\end{equation}

From \eqref{diseq:D1} and \eqref{diseq:D2} it stems  that

$$K_{1}-K_{2}= \left[\frac{(\gamma^{'})^2- \gamma^{''}\gamma}{\gamma^2} \right]- \left[\frac{(\alpha^{'})^2- \alpha^{''}\alpha}{\alpha^2} \right]= D_{\gamma}-D_{\alpha}$$

with $D_{\gamma}-D_{\alpha}>0$ for assumption (see Table~ \ref{tab:01}). In what follows we abbreviate $D_{\varphi}$ as $D$. Hence we have found the following conditions

\begin{equation}
\begin{cases}
\mathbf{A} \ D \le D_{\alpha}+ K_{1}\\
\mathbf{B} \ D \le D_{\gamma}+ K_{2}\\
\end{cases}
\label{diseq:system}
\end{equation}

Condition $\mathbf{B}$ implies $D \le D_{\gamma}+ K_{2}+(K_{1}-K_{2})=D_{\gamma}+K_{1}$ and yields the further condition

$$\mathbf{C} \ D \le D_{\gamma}+K_{1}$$

Thus, we may collect the three conditions already found

\begin{equation}
\begin{cases}
\mathbf{A} \ D \le D_{\alpha}+ K_{1}\\
\mathbf{B} \ D \le D_{\gamma}+ K_{2}\\
\mathbf{C} \ D \le D_{\gamma}+K_{1}\\
\end{cases}
\label{diseq:system2}
\end{equation}

Now we may distinguish three separate cases which satisfy the condition $K_{1}-K_{2}>0$:\\

(a) $ K_{1},\ K_{2}>0$

We use conditions $\mathbf{B}, \mathbf{C}$
\begin{equation}
\begin{cases}
\mathbf{B} \ D \le D_{\gamma}+ K_{2}\\
\mathbf{C} \ D \le D_{\gamma}+K_{1}\\
\end{cases} \rightarrow 
\begin{cases}
  2D \le 2D_{\gamma}+ 2K_{2}\\
  D \le D_{\gamma}+2K_{1}\\
\end{cases} \rightarrow
\begin{cases}
  D \le D_{\gamma}+2(K_{2}-K_{1})=D_{\gamma}\\
  -\\
\end{cases}
\label{diseq:system3}
\end{equation}

and we conclude that  $D \le D_{\gamma}$.\\

(b) $ K_{1}>0,\ K_{2}<0$

By applying condition $\mathbf{B}$ , it follows  $D \le D_{\gamma}$.\\

(c) $ K_{1}<0,\ K_{2}<0$

By applying condition $\mathbf{B}$ or $\mathbf{C}$  , it follows  $D \le D_{\gamma}$.\\

We have proved that for any possible sign of $ K_{1}, \ K_{2}$, $D \le D_{\gamma}$. By definition of effective sample size from \cite{morita2008determining} we know that

\begin{align*}
ESS(\varphi(\theta))&=\underset{m \in \mathbb{N}}{\mbox{Argmin}}\{\delta(m, \bar{\theta}, \varphi, q_{m}) \}=\\
&=\underset{m \in \mathbb{N}}{\mbox{Argmin}} \{  | D
-D_{q_{m}}(\bar{\theta})|  \} \\
\end{align*}

evaluated in the plug-in estimate $\bar{\theta}=E_{\pi}[\theta]$. From Table~\ref{tab:01} we also know that the observed information of the baseline posterior $D_{q_{m}}$  is a linear function of the sample size $m$ and is increasing: $$\frac{d D_{q_{m}} }{dm}>0, \ \forall m \in \mathbb{N}$$ Thus we may conclude that from $D \le D_{\pi}$ it follows:

\begin{align*}
 ESS(\varphi(\theta))&= \underset{m \in \mathbb{N}}{\mbox{Argmin}} \{  | D
_{\varphi}(\bar{\theta})
-D_{q_{m}(\theta|y)}(\bar{\theta})|  \} \le \\
&\le  \underset{m \in \mathbb{N}}{\mbox{Argmin}} \{  | D
-D_{q_{m}}(\bar{\theta})|  \} = ESS(\pi(\theta))\ . \ \  \Box
\end{align*}

\end{small}

\newpage

\subsection*{Logistic regression for phase I trial}
\vspace{0.5cm}

\begin{small}

\textit{Algorithm for computing the ESS} \citep{morita2008determining}
\vspace{0.5cm}

\begin{itemize}
\item According to the definitions in \eqref{eq:morita:derivative}, we compute the following quantities:\\ $D_{\pi,1}=(\tilde{\sigma}^{2}_{\mu})^{-1},\ D_{\pi,2}=(\tilde{\sigma}^{2}_{\beta})^{-1}$.
\item We need to compute $D_{q,1}(m,\bm{\theta}, X_{m}, \bm{y}_{m})=\sum_{i=1}^{m} \pi(X_{i}, \theta)\{ 1-\pi(X_{i}, \theta) \}$,\\ $ D_{q,2}(m,\bm{\theta}, X_{m}, \bm{y}_{m})=\sum_{i=1}^{m}X^{2}_{i} \pi(X_{i}, \theta)\{ 1-\pi(X_{i}, \theta) \} $.
\item  It turns out that $\int D_{q_{m},j}f(\bm{y}_{m})d\bm{y}_{m}$ ---where $f(\bm{y}_{m})$ is the likelihood \eqref{eq:likelihood_logistic} evaluated in correspondence of fixed values for $\bm{\theta}$ and $\bm{X}$--- cannot be computed analytically and need to be computed through Monte Carlo simulation. Before of proceeding, let us notice that $D_{q,1}(m,\bm{\theta}, X_{m}, \bm{y}_{m})$ and $D_{q,2}(m,\bm{\theta}, X_{m}, \bm{y}_{m})$ depend on $X_{m}$ but not on $\bm{y}_{m}$, and this simplifies the simulation procedure. We may replace them respectively with the new notations $D_{q,1}(m,\bm{\theta}, X_{m})$ and $D_{q,2}(m,\bm{\theta}, X_{m})$.
\item Assuming a uniform distribution for the doses, we draw $X^{(t)}_{1},...,X^{(t)}_{6}$ independently from $\{X_{1},...,X_{6} \}$ with probability 1/6 each, for $t=1,...,100000$.
\item Use the Monte Carlo average $T^{-1} \sum_{t=1}^{T}D_{q,j}(m,\bm{\theta}, X_{m})$ in place of $\int D_{q_{m},j}f(\bm{y}_{m})d\bm{y}_{m}$, for $ j=1,2$.
\item  Compute $\delta_{1}(m_{\mu},\bar{\bm{\theta}}, \pi_{\mu}, q_{m_{\mu}})$, $ \delta_{2}(m_{\beta},\bar{\bm{\theta}}, \pi_{\beta}, q_{m_{\beta}})$ and $ \delta(m,\bar{\bm{\theta}}, \pi, q_{m})$.
\item $ ESS(\pi(\mu)), ESS(\pi(\beta))$ and $ESS$ are the interpolated values of the sample sizes $m_{\mu},  m_{\beta},  m$ minimizing $\delta_{1}, \delta_{2}$ and $\delta$ respectively.
\end{itemize}

\newpage

\begin{table}
\caption{\label{tab:03}Effective sample sizes $ESS(\pi(\mu)), ESS(\pi(\beta))$ for the tolerable dose in a phase I trial.}
\begin{tabular}{|llll|} 
\hline
& \tiny{$ ESS$} & \tiny{$ 
  ESS(\pi(\mu))$}& \tiny{$ESS(\pi(\beta))$}\\
  \hline
  $\sigma^{2}_{\mu}=\sigma^{2}_{\beta}=0.5^{2}$ & 37.00 & 22.73 & 98.11 \\
  $\sigma^{2}_{\mu}=\sigma^{2}_{\beta}=1^{2}$ & 10.00 & 5.75 & 25.56 \\
  $\sigma^{2}_{\mu}=\sigma^{2}_{\beta}=2^{2}$ & 3.00 & 1.37 & 6.53 \\
 $\sigma^{2}_{\mu}=\sigma^{2}_{\beta}=3^{2}$ & 2.00 & 1.03 & 3.06 \\
 $\sigma^{2}_{\mu}=\sigma^{2}_{\beta}=5^{2}$ & 1.00 & 1.00 & 1.38 \\
 \hline
\end{tabular}
\end{table}

\begin{table}
\caption{\label{tab:04} \small Effective sample sizes $ESS(\varphi(\mu)), \ ESS(\varphi(\beta))$ for the MDD priors $\varphi(\mu)=\psi \mathcal{N}(\tilde{\mu}_{\mu}, c\tilde{\sigma}^{2}_{\mu})+ (1-\psi)\mathcal{N}(\tilde{\mu}_{\mu}, \tilde{\sigma}^{2}_{\mu}), \ \varphi(\beta)=\psi \mathcal{N}(\tilde{\mu}_{\beta}, c\tilde{\sigma}^{2}_{\beta})+ (1-\psi)\mathcal{N}(\tilde{\mu}_{\beta}, \tilde{\sigma}^{2}_{\beta})$ according to different values of the mixture weight $\psi$.}
\begin{small}

\begin{tabular}{|l|lll|lll|lll|}
\hline
 &  \multicolumn{3}{|c|}{$\psi=0.2$}&  \multicolumn{3}{|c|}{$\psi=0.5$}&\multicolumn{3}{|c|}{$\psi=0.8$}\\
   & \tiny{$ 
  ESS$} & \tiny{$ ESS(\varphi(\mu))$}& 
  \tiny{$ESS(\varphi(\beta))$} & \tiny{$ ESS
  $} & \tiny{$ ESS(\varphi(\mu))$}& \tiny{$ESS(\varphi(\beta))$}& 
  \tiny{$ ESS$} & \tiny{$ ESS(\varphi(\mu))
  $}& \tiny{$ESS(\varphi(\beta))$} \\ 
  \hline
$\sigma^{2}_{\mu}=\sigma^{2}_{\beta}=0.5^{2}$ &  37.00 & 22.70 & 98.06 & 37.00 & 22.62 & 97.90 & 37.00 & 22.30 & 97.18   \\ 
  $\sigma^{2}_{\mu}=\sigma^{2}_{\beta}=1^{2}$ &  10.00 & 5.73 & 25.50 & 10.00 & 5.69 & 25.31 & 9.00 & 5.52 & 24.58   \\ 
  $\sigma^{2}_{\mu}=\sigma^{2}_{\beta}=2^{2}$ &  3.00 & 1.37 & 6.49 & 3.00 & 1.37 & 6.42 & 3.00 & 1.31 & 6.06   \\ 
  $\sigma^{2}_{\mu}=\sigma^{2}_{\beta}=3^{2}$  & 2.00 & 1.03 & 3.03 & 2.00 & 1.03 & 3.01 & 2.00 & 1.03 & 2.68   \\ 
  $\sigma^{2}_{\mu}=\sigma^{2}_{\beta}=5^{2}$  & 1.00 & 1.00 & 1.38 & 1.00 & 1.00 & 1.37 & 1.00 & 1.00 & 1.26  \\ 
   \hline
\end{tabular}
\end{small}
\end{table}

\begin{table}
\caption{\label{tab:05} \small Effective sample sizes $ESS(\varphi(\mu)), \ ESS(\varphi(\beta))$ for the MDD priors $\varphi(\mu)=\psi+(1-\psi)\pi_{\mu}, \ \varphi(\beta)=\psi+(1-\psi)\pi_{\beta}$ according to different values of the mixture weight $\psi$.}
\begin{small}
\begin{tabular}{|l|lll|lll|lll|}
\hline
 &  \multicolumn{3}{|c|}{$\psi=0.2$}&  \multicolumn{3}{|c|}{$\psi=0.5$}&\multicolumn{3}{|c|}{$\psi=0.8$}\\
   & \tiny{$ 
  ESS$} & \tiny{$ ESS(\varphi(\mu))$}& 
  \tiny{$ESS( \varphi(\beta))$} & \tiny{$ ESS
  $} & \tiny{$ ESS(\varphi(\mu))$}& \tiny{$ESS(\varphi(\beta))$}& 
  \tiny{$ ESS$} & \tiny{$ ESS(\varphi(\mu))
  $}& \tiny{$ESS(\varphi(\beta))$} \\ 
  \hline
  $\sigma^{2}_{\mu}=\sigma^{2}_{\beta}=0.5^{2}$ & 32.00 & 19.71 & 87.65 & 23.00 & 14.03 & 62.43 & 11.00 & 6.55 & 29.06 \\
  $\sigma^{2}_{\mu}=\sigma^{2}_{\beta}=1^{2}$ & 6.00 & 3.58 & 15.78 & 3.00 & 1.68 & 7.42 & 1.00 & 1.03 & 2.48 \\ 
  $\sigma^{2}_{\mu}=\sigma^{2}_{\beta}=2^{2}$ & 1.00 & 1.00 & 1.99 & 1.00 & 1.00 & 1.14 & 1.00 & 1.00 & 1.03 \\
  $\sigma^{2}_{\mu}=\sigma^{2}_{\beta}=3^{2}$ &  1.00 & 1.00 & 1.10 & 1.00 & 1.00 & 1.03 & 1.00 & 1.00 & 1.03 \\
  $\sigma^{2}_{\mu}=\sigma^{2}_{\beta}=3^{2}$ & 1.00 & 1.00 & 1.03 & 1.00 & 1.00 & 1.03 & 1.00 & 1.00 & 1.03 \\ 
   \hline
\end{tabular}
\end{small}
\end{table}

\end{small}

\end{document}